\documentclass[12pt,preprint]{aastex}
\usepackage{graphics}
\usepackage{graphicx}
\usepackage{epsfig}

\newcommand{\beq}{\begin{equation}}
\newcommand{\eeq}{\end{equation}}
\newcommand{\beqa}{\begin{eqnarray}}
\newcommand{\eeqa}{\end{eqnarray}}
\newcommand{\bit}{\begin{itemize}}
\newcommand{\eit}{\end{itemize}}

\begin{document}

\title{Time evolution of Wouthuysen-Field coupling}

\author{Ishani Roy\altaffilmark{1}, Wen Xu\altaffilmark{2},
Jing-Mei Qiu\altaffilmark{3}, Chi-Wang
Shu\altaffilmark{1}, and Li-Zhi Fang\altaffilmark{2}}

\altaffiltext{1}{Division of Applied Mathematics, Brown University,
Providence, RI 02912}
\altaffiltext{2}{Department of Physics,
University of Arizona, Tucson, AZ 85721}
\altaffiltext{3}{Department of Mathematical and Computer Science,
Colorado School of Mines, Golden, CO 80401}

\begin{abstract}

We study the Wouthuysen-Field coupling at early universe with
numerical solutions of the integrodifferential equation describing
the kinetics of photons undergoing resonant scattering. The
numerical solver is developed based on the weighted essentially
non-oscillatory (WENO) scheme for the Boltzmann-like
integrodifferential equation. This method has perfectly passed the
tests of analytic solution and conservation property of the resonant
scattering equation. We focus on the time evolution of the
Wouthuysen-Field (W-F) coupling in relation to the 21 cm emission
and absorption at the epoch of reionization. We especially pay
attention to the formation of the local Boltzmann distribution,
$e^{-(\nu-\nu_0)/kT}$, of photon frequency spectrum around resonant
frequency $\nu_0$ within width $\nu_l$, i.e. $|\nu-\nu_0| \leq
\nu_l$. We show that a local Boltzmann distribution will be formed
if photons with frequency $\sim \nu_0$ have undergone a ten thousand
or more times of scattering, which corresponds to the order of
10$^3$ yrs for neutral hydrogen density of the concordance
$\Lambda$CDM model. The time evolution of the shape and width of the
local Boltzmann distribution actually doesn't dependent on the
details of atomic recoil, photon sources, or initial conditions very much.
However, the intensity of photon flux at the local Boltzmann
distribution is substantially time-dependent. The time scale of
approaching the saturated intensity can be as long as 10$^{5}$-$10^{6}$
yrs for typical parameters of the $\Lambda$CDM model. The intensity
of the local Boltzmann distribution at time less than 10$^{5}$ yrs is
significantly lower than that of the saturation state. Therefore, it
may not be always reasonable to assume that the deviation of the
spin temperature of 21 cm energy states from cosmic background
temperature is mainly due to the W-F coupling {\it if} first stars
or their emission/absorption regions evolved with a time scale equal
to or less than Myrs.

\end{abstract}

\keywords{cosmology: theory - intergalactic medium - radiation
transfer - scattering}

\section{Introduction}

It is generally believed that the physical state of the universe at
the epoch of reionization can be probed by detecting the redshifted
21 cm signals from the ionized and heated regions around the first
generation of stars (e.g. Furlanetto, Oh, \& Briggs 2006).  The
emission and absorption of 21 cm are caused by the deviation of the
spin temperature $T_s$ of neutral hydrogen from the temperature of
cosmic microwave background (CMB) $T_{\rm CMB}$ at the considered
redshift $z$. Many calculations have been done on  the 21 cm
emission and absorption from ionized halos of the first stars
(Chuzhoy et al. 2006; Cen 2006; Liu et al. 2007). A common
assumption of these calculations is that the deviations of $T_s$
from $T_{\rm CMB}$ are mainly due to the Wouthuysen-Field (W-F)
coupling (Wouthuysen, 1952; Field, 1958, 1959). That is, the
resonant scattering of Ly$\alpha$ photons with neutral hydrogen
atoms locks the color temperature $T_c$ of the photon spectrum
around the Ly$\alpha$ frequency to be equal to the kinetic
temperature of hydrogen gas $T$. Consequently, the spin degree of
freedom is determined by the kinetic temperature of hydrogen gas
$T$.

The W-F coupling is from the kinetics of photons undergoing resonant
scattering, which is described by a Boltzmann integrodifferential
equation. All the above-mentioned calculations are based on
time-independent solution of the resonant scattering kinetic equation with
Fokker-Planck approximation (Chen \& Miralda-Escude, 2004; Hirata, 2006; Furlanetto \&
Pritchard 2006; Chuzhoy \& Shapiro 2006). This is equal to assume that
the time scale of the onset of the W-F coupling is less than all time scales
related to the 21 cm emission/absorption. However, even in the first paper
of the W-F coupling, the problem of time scale has been addressed as
follows: ``One can infer from this fact that the photons (in a box),
after an infinite number of scattering processes on gas atoms with
kinetic temperature $T$, will obtain a statistical distribution over the spectrum
proportional to the Planck-radiation spectrum of temperature $T$.
After a finite but large number of scattering processes, the Planck
shape will be produced in a region around the initial frequency''
(Wouthuysen, 1952). That is, the W-F coupling is onset only ``after a
finite but large number of scattering''. One cannot assume
that the time-independent solution is available for the W-F coupling if
the time-scales of the evolution of the first stars and their
emission/absorption regions are short. A study on the time evolution
of radiation spectrum due to resonant scattering is necessary.

This problem is especially important for the 21 cm signal from the
first stars, as the life times of the first stars are short. The
ionized and heated regions around the first stars are strongly
time-dependent (Cen 2006; Liu et al. 2007). The 21 cm emission/absortion
regions are located in a narrow shell just outside the ionized region. On the
other hand, the speed of the ionization-front (I-front) is rather
high, even comparable to the speed of light. The time scales of the
formation and the evolution of the 21 cm regions can be estimated by
$\sim d/c$, $d$ being the thickness of the shells of 21 cm emission
and absorption.  The time-independent solution would be proper only
if Ly$\alpha$ photons approach the time-independent state in a time
shorter than that of the 21 cm region evolution.

Very few works have been done on the time dependent behavior of the
W-F coupling. There is the lack of a time dependent solution even
for the Fokker-Planck approximation. The existed time-dependent
solvers (e.g. Meiksin, 2006) cannot pass the tests of analytical
solutions (Field, 1959). On the other hand, the WENO algorithm is
found to be effective to solve the Boltzmann equation (Carrillo et
al. 2003) and radiative transfer (Qiu et al. 2006, 2007, 2008). In
this paper, we will study the time-dependent behavior of the W-F
coupling with the WENO method by numerically solving the
integrodifferential equation. In this context, we also develop a
numerical solver in accordance with the term of resonant scattering.

The paper is organized as follows. Section 2 presents the basic equations of
the resonant scattering of photons. Section 3 very briefly mentions the numerical
solver of the WENO scheme and its test with Field's analytic solutions, leaving
the details of algorithmic issues to the Appendix. Section 4
presents the time-dependent W-F coupling with a static background. Section 5 shows the
numerical results of the W-F coupling in an expanding background. Finally, conclusions
are given in Section 6, in which the application to the 21 cm problem is also
addressed.

\section{Basic equations}

\subsection{Radiative transfer equations with resonant scattering}

Since we focus on the time-scales of the W-F coupling, we consider a
homogeneous and isotropically expanding infinite medium consisting
of neutral hydrogen with temperature $T$. The kinetics of photons in
the frequency space is described by the radiative transfer equation
with resonant scattering (Hummer \& Rybicki, 1992; Rybicki \&
Dell'antonio 1994)
\begin{eqnarray}
\label{eq1}
 & & \frac{\partial J(x, t)}{\partial t}+2HJ(x, t)-
\frac{cH}{v_T}\frac{\partial J(x, t)}{\partial x}=  \nonumber\\
   & & -kc \phi(x)J(x, t)
+kc\int \mathcal{R}(x,x')J(x',t)dx'+S(x,t)
\end{eqnarray}
where $J$ is the flux in terms of the photon number in units
s$^{-1}$cm$^{-2}$. $H(t)=\dot{a}(t)/a(t)$ is the Hubble parameter,
$a(t)$ being the cosmic factor; $v_T=(2k_BT/m)^{1/2}$ is the thermal
velocity of hydrogen atoms. The dimensionless frequency $x$ is
related to the frequency $\nu$ and the resonant frequency $\nu_0$ by
$x = (\nu-\nu_0)/\Delta \nu_D$, and $\Delta \nu_D=\nu_0v_T/c$ is the
Doppler broadening. $S(t,x)$ is the source of photons. The parameter
$k=\chi/\Delta \nu_D$, where $\chi$ is the intensity of the resonant absorption
given by $\chi=\pi e^2n_1f_{12}/m_ec$, and $n_1$ being the
number density of neutral hydrogen HI at ground state,
$f_{12}=0.416$ is the oscillator strength. The cross section of resonant
scattering at the line center is
\begin{equation}
\sigma_0=\frac{\pi e^2}{m_ec}f_{12}(\Delta \nu_D)^{-1}.
\end{equation}
In eq.(\ref{eq1}), $\phi(x)$ is the profile of the absorption line
at the resonant frequency $\nu_0$. If the profile is dominated by
Doppler broadening, we have
\begin{equation}
\label{eq3}
\phi(x) = \frac{1}{\sqrt{\pi}}e^{-x^2} .
\end{equation}
The redistribution function $\mathcal{R}(x,x')$ of eq.(\ref{eq1}) gives the
probability of a photon absorbed at frequency $x'$, and
isotropically re-emitted at frequency $x$. For coherent
scattering, we have (Field 1958; Hummer 1962; Basko 1981)
\begin{equation}
\label{eq:R}
\mathcal{R}(x,x')=\frac{1}{2}e^{2bx'+b^2}{\rm
erfc}[{\rm max}(|x+b|,|x'+b|)],
\end{equation}
where the parameter $b=h\nu_0/mv_T c= 2.5\times 10^{-4} (10^4/T)^{1/2}$
is due to the recoil of atoms. It is in the range of $3\times
10^{-2}$ - $3\times 10^{-4}$, if the temperature $T$ is in the range of $1$ K
- $10^4$ K. The redistribution function is normalized as
\begin{equation}
\int \mathcal{R}(x,x') dx= \phi(x').
\end{equation}
Therefore, we have
\begin{equation}
\label{eq6}
kc \int \phi(x)J(x,t)dx= kc \int\int \mathcal{R}(x,x')J(x',t) dxdx'.
\end{equation}
It means that the total number of photons absorbed given by the term
$kc\phi(x)J(x,t)$ of eq.(\ref{eq1}) is equal to the total number of
scattered photons.  Therefore, with eq.(\ref{eq1}), the number of
photons is conserved.

\subsection{Rescaling the equations}

We use the new time variable $\tau$ defined as $\tau= cn_1\sigma_0
t$, which is in unit of the mean free flight time of photons at
resonant frequency. For the concordance $\Lambda$CDM model, the
number density of hydrogen atoms is $n_{\rm H}= 1.88\times 10^{-7}
(\Omega_bh^2/0.022)(1+z)^3$ cm$^{-3}$. Therefore, we have
\begin{equation}
 t=0.054 f^{-1}_{\rm
HI}\left(\frac{T}{10^4}\right)^{1/2}\left (\frac{10}{1+z}\right
)^3\left (\frac{0.022}{\Omega_b h^2}\right )\tau \hspace{3mm} yrs,
\end{equation}
where $f_{\rm HI}=n_1/n_{\rm H}$ is the fraction of neutral
hydrogen.

We rescale the eq.(\ref{eq1}) by the following new variables
\begin{equation}
 J'(x,\tau)= [a(t)/a(0)]^2J(x,t), \hspace{5mm} S'(x,\tau)=
[a(t)/a(0)]^2S(x,t).
\end{equation}
Thus, eq.(\ref{eq1}) becomes
\begin{eqnarray}
\label{eq9}
\frac{\partial J'(x, \tau)}{\partial \tau}& =& - \phi(x)J'(x, \tau) \nonumber \\
 & & +\int \mathcal{R}(x,x')J'(x',\tau)dx' + \gamma \frac{\partial J'}{\partial x}
+S'(x,\tau),
\end{eqnarray}
where the parameter $\gamma$ is the so-called Sobolev parameter.
$\gamma= (H/v_T k)=(8\pi H/3A_{12}\lambda^3 n_1)=(H m_e\nu_0/\pi
e^2n_1f_{12})$, where $\lambda$ is the wavelength for Ly$\alpha$
transition. $\gamma^{-1}$ measures the number of scattering during a
Hubble time. It actually is the Gunn-Peterson optical depth given by
by
\begin{equation}
\label{eq10} \gamma^{-1}= 4.9 \times 10^5 h^{-1}f_{\rm
HI}\left(\frac{0.25}{\Omega_M}\right
)^{1/2}\left(\frac{\Omega_bh^{2}}{0.022}\right )
\left(\frac{1+z}{10}\right )^{3/2}.
\end{equation}
Around the first stars, the number $f_{\rm HI}$ is strongly
dependent on time and position (Liu et al. 2007). It is as small as
$10^{-6}$ within the ionized sphere, and as high as $\simeq1$
outside the ionized sphere. Therefore, the parameter $\gamma$ would
be in the range from 1 to $10^{-6}$.

The physical meaning of the terms on the right hand side of
eq.(\ref{eq9}) is clear. The first term is the absorption at
frequency $x$, the second term is the re-emission of photons with
frequency $x$ by scattering, and the third term describes the
redshift of photons. The time scale of a photon moving $\Delta x$ in
the frequency space is equal to
\begin{equation}
\label{eq11}
\Delta \tau = \gamma^{-1}\Delta x .
\end{equation}
This actually is due to the Hubble expansion.

Considering eq.(\ref{eq6}), eq.(\ref{eq9}) gives
\begin{equation}
\label{eq12}
\frac{d}{d\tau}\int J' dx = \int S' dx.
\end{equation}
This equation shows that the total number of photons
$\displaystyle\int J' dx$ is dependent only on the sources,
regardless of the parameter $b$ of the resonant scattering. Since
numerical errors accumulated over a long time evolution could be huge,
Eq.(\ref{eq12}) is useful to check the reliability of a numerical
code. We will use $J$ for $J'$ and $S$ for $S'$ in sections below.
It will not cause confusion.

\section{Numerical method}

We use the WENO scheme to solve the eq.(\ref{eq9}). This algorithm
has been given in Roy et al. (2009). Some of the algorithmic details
are given in the Appendix. We
present a test to show the good performance of our solver below.

Figure 1 plots both the analytical (Field, 1959) and WENO numerical
solutions of eq.(\ref{eq9}) with parameters $\gamma=0$ and $b=0$. It
shows that the numerical solutions can follow the analytical
solution in all the time $t$ and the frequency $x$ considered. This
result is not trivial if compared with the results of other
numerical solvers, such as Meiksin (2006), which shows a large
deviation between the analytical and numerical solutions. Therefore,
our scheme is more reliable.
\begin{figure}[htb]
\centering
\includegraphics[width=6.5cm]{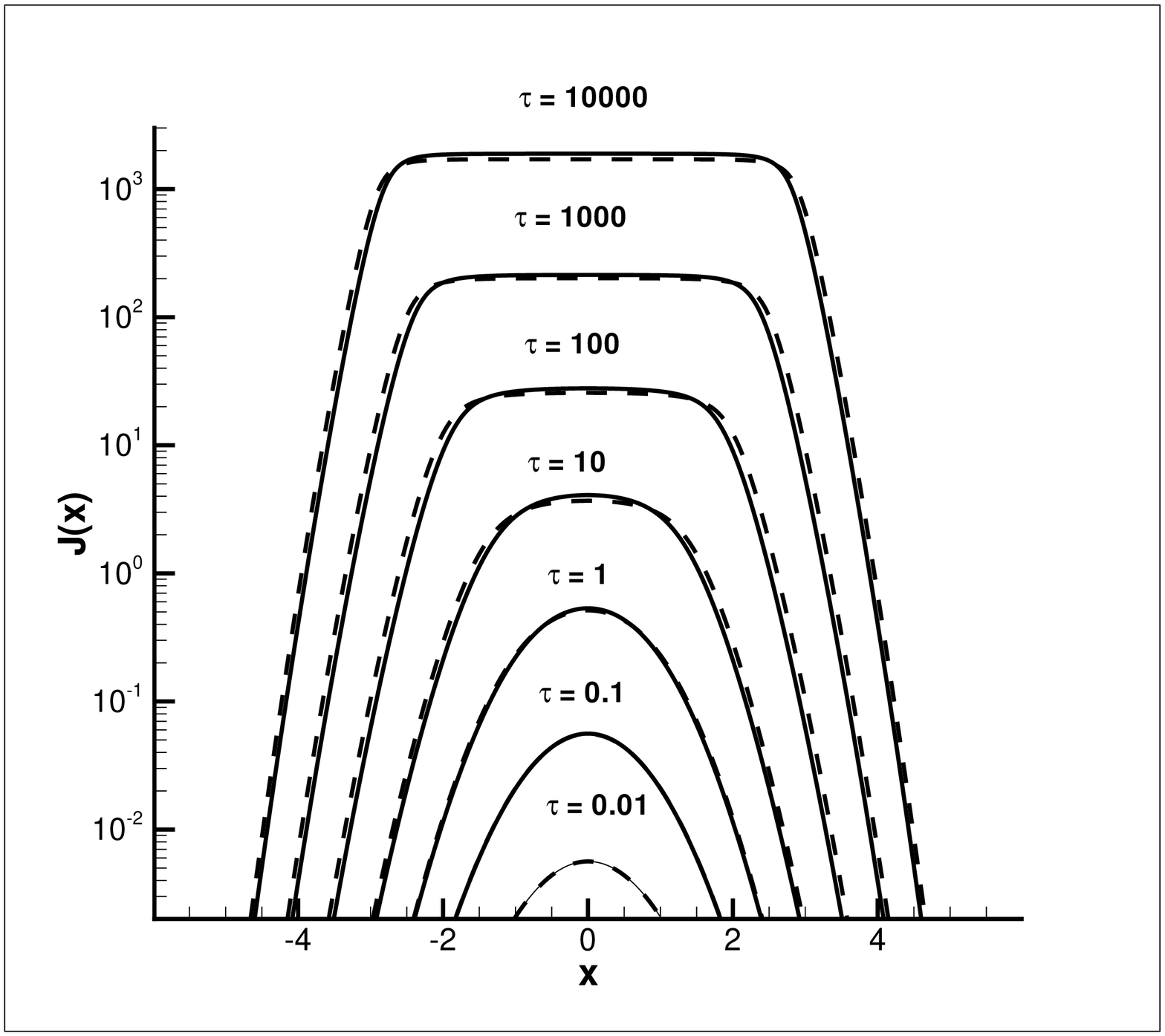}
\caption{Analytical (dashed) and WENO numerical (solid) solutions
of eq.(\ref{eq9}) with $\gamma=0$ and $b=0$. The source is taken to
be $S=\phi(x)$ and
initial condition $J(x,0)=0$.
}
\label{fig1}
\end{figure}

\section{Wouthuysen-Field coupling in a static background}

To study the effect of atomic recoil, we first solve the time
evolution of $J(x,\tau)$ in a static background, i.e. $\gamma=0$. A
typical time-dependent result is shown in Figure 2, in which
$b=0.03$. The solution of Figure 2 is actually similar to that shown
in Figure 1. Figure 1 shows that a flat plateau around $x=0$ is to
be formed at $\tau>100$, while Figure 2 shows that $J(x,\tau)$
evolves into a Boltzmann distribution around $x=0$ as
\begin{equation}
\label{eq13} J(x,\tau) \simeq J(0,\tau)e^{-2bx}=J(0,\tau)e^{-h(\nu -
\nu_0)/kT}, \hspace{5mm}  |x| \leq x_l.
\end{equation}
This is the so-called ``Planck shape in a region around the initial
frequency'' (Wouthuysen, 1952). The expression eq.(\ref{eq13}) has
also been found by Field (1959). We will call this feature to be a
local Boltzmann distribution. The width $x_l$ of the local Boltzmann
distribution is numerically defined by the frequency range $|x| \leq
x_l$, in which the slope $\ln J(x,\tau)/d x$ deviating from $2b$ is
small (see below).
\begin{figure}[htb]
\centering
\includegraphics[width=7.5cm]{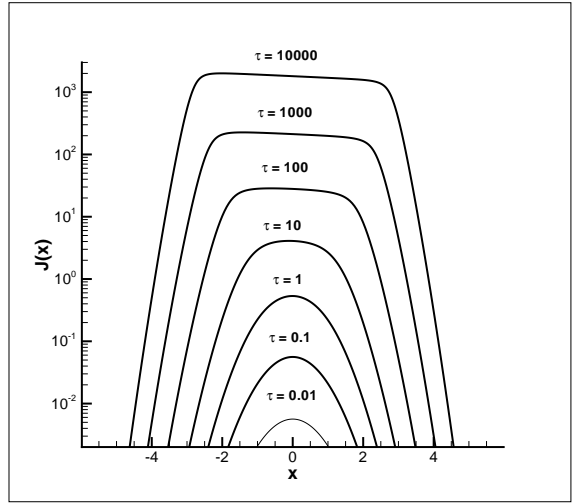}
\caption{WENO numerical solutions of eq.(\ref{eq9}) with $\gamma=0$
and $b=0.03$. The source
is taken to be $S=\phi(x)$ and initial condition $J(x,0)=0$.
}
\label{fig2}
\end{figure}

 Figure 3 plots the time-evolution of $J(x,\tau)$ with different
parameter $b$, and a zoom-in figure at time $\tau=10^4$. The zoom-in
figure shows clearly that the integral $\int J(x,\tau)dx$ is $b$-independent.
This is consistent with the photon number conservation eq.(\ref{eq12}). It
shows again that the WENO algorithm is robust. We can also see from
the right panel of Figure 3 that all the curves of $\ln J(x,\tau)$ vs. $x$
at $\tau=10^4$ and in the range $-2<x <2$ can be approximated as a straight
line. That is, the width $x_l$ of the local Boltzmann distribution shown
in Figure 3 is equaal to about 2, and it is approximately $b$-independent.

The formation and evolution of the local Boltzmann distribution
can be quantitatively described by $B(\tau)$ defined as
\begin{equation}
B(\tau)=2b/[\ln J(0,\tau) - \ln J(1,\tau)],
\end{equation}
where $\ln J(0,\tau) - \ln J(1,\tau)$ is the slope of the straight
line $\ln J(x,\tau)$ vs. $x$ for $|x| \leq 1$.  For Gaussian
source $S(x)=\phi(x)=e^{-x^2}/\sqrt{\pi}$, we have $B(0)=2b$, and
$B(\tau)$ approaches $2b$ at large $\tau$. Figure 4 presents the
numerical relation of $B(\tau)$ vs. $\tau$. The slopes
$[\log J(0,\tau) - \log J(1,\tau)]$ at $\tau=10^5$ are,
respectively, $0.0601$ for $b = 0.030$, $0.0303$ for $b=0.015$,
$0.0159$ for $b = 0.0079$, and $0.0051$ for $b=0.0025$. That is,
within the frequency range $|x| \leq x_l$ and $x_l=1$, the
relative deviation of the slope $d \ln J(x,\tau)/d x$ from $2b$ is no
larger than 2\%. Thus, $\tau=10^{5}$ can be considered as the time scale
of forming a local Boltzmann distribution within $|x|<x_l=1$. For small
width $x_l<1$, this time scale is lower as $\simeq 10^4$. Therefore,
the time scale of the onset of W-F coupling with
the width $x_l$ equal to about Doppler broadening is $10^{4}$-$10^{5}$.

\begin{figure}[htb]
\centering
\includegraphics[width=7.5cm]{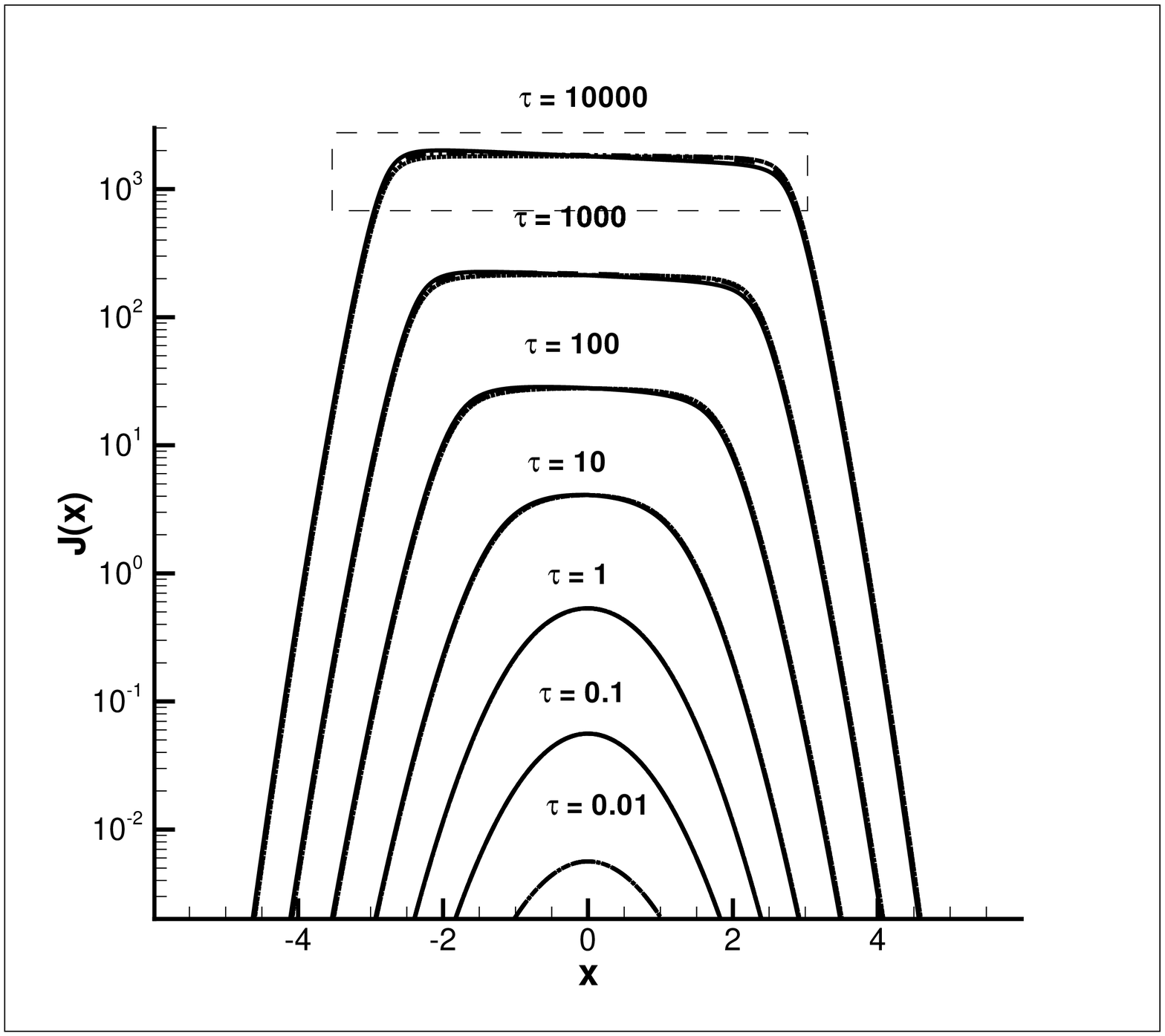}
\includegraphics[width=7.5cm]{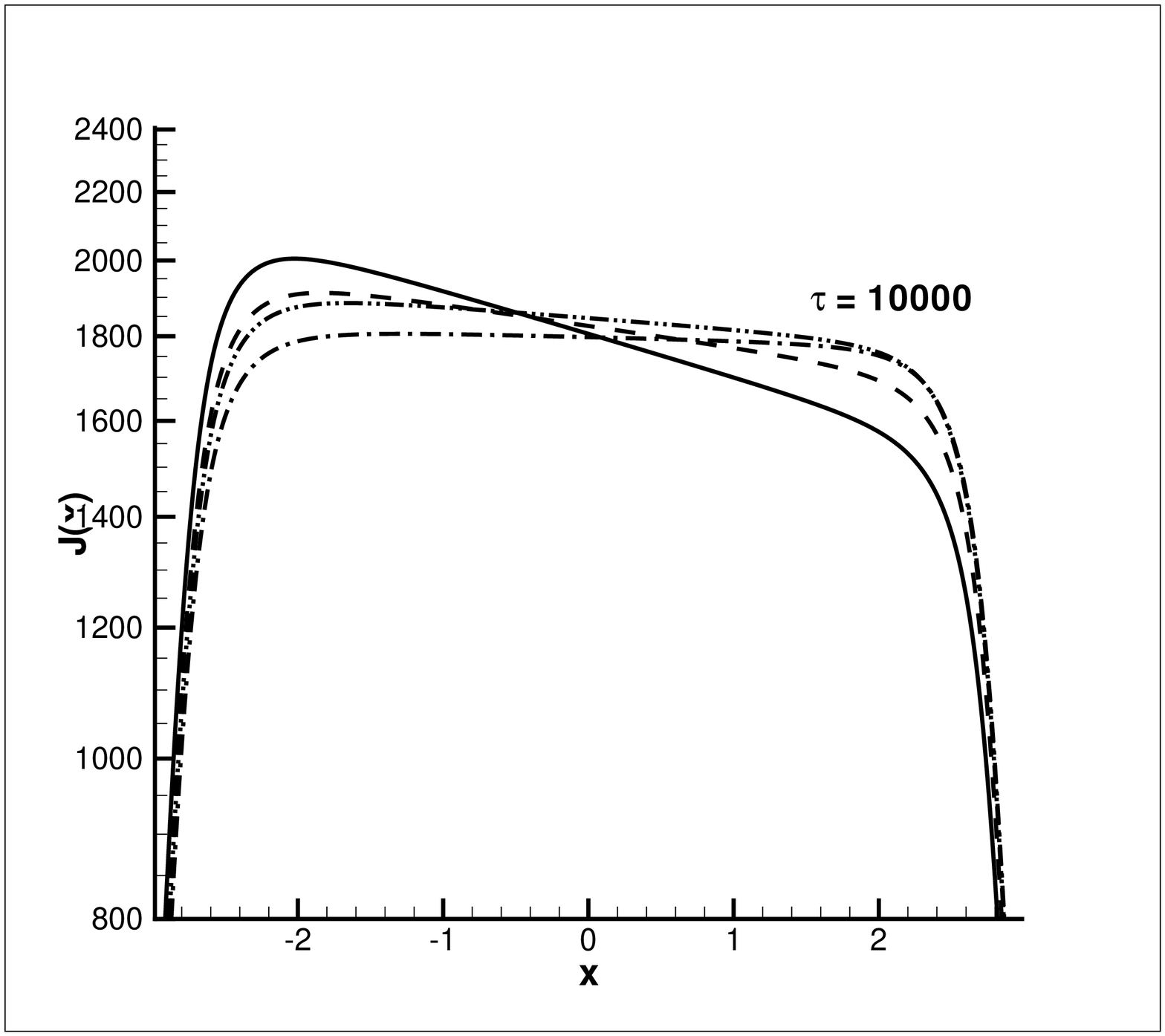}
\caption{WENO numerical solutions of eq.(\ref{eq9}) with $\gamma=0$ and
$b=0.03$ (solid), 0.015 (dashed), 0.0079 (dot-dot-dashed) and 0.0025
(dot-dashed). The source is taken to be $S=\phi(x)$ and initial
condition $J(x,0)=0$. The right panel is a zoom-in of the dashed
square of the left panel.} \label{fig3}
\end{figure}

\begin{figure}[htb]
\centering
\includegraphics[width=7.5cm]{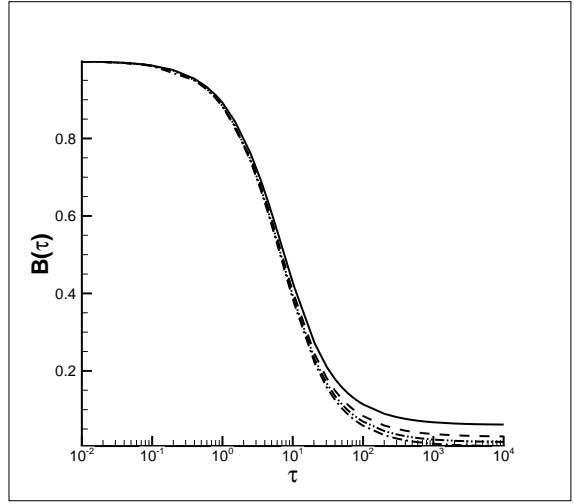}
\caption{$B(\tau)$ vs. $\tau$ for solutions of Figure 3 with
$b=0.03$ (solid), 0.015 (dashed), 0.0079 (dot-dot-dashed) and 0.0025
(dot-dashed).} \label{fig4}
\end{figure}

We can relate the width $x_l$ with the mean number of scattering,
$N_c$, needed to form the local Boltzmann distribution.  Although
the redistribution function eq.(\ref{eq:R}) is $b$-dependent, the
probability of $x$ photons undergoing a resonant scattering per unit
time is $\phi(x)$, which is $b$-independent. Thus, at a given time
$\tau$, the mean number $N_c$ of resonant scattering of photons  
within $|x|\leq x_l$ approximately is
\begin{equation}
\label{eq15} N_c\simeq  \tau \frac{1}{x_l}\int_{0}^{x_l} \phi(x)dx.
\end{equation}
Eq.(15) gives the ``finite but large number of scattering'' for
realizing a local Boltzmann distribution within $x \leq x_l(\tau)$
(Wouthuysen, 1952). Therefore, the approximate $b$-independence of
$x_l$ (Figures 3 and 4) would imply the $b$-independence of $N_c$.

\section{Wouthuysen-Field coupling in an expanding background}

\subsection{Width of the local Boltzmann distribution}

Considering an expanding background, i.e. $\gamma\neq 0$, we solve
eq.(9) by the WENO algorithm. Figure 5 plots solutions with the same
source $S=\phi(x)$ and parameter $b=0.03$ as in Figure 2, but with
$\gamma=10^{-3}$ and $10^{-5}$. Similar to Figure 2, a local
Boltzmann distribution has formed when $\tau
\geq  10^3$ for both $\gamma=10^{-3}$ and $10^{-5}$. The section of
the spectrum near $x=0$ becomes $\tau$-independent when $\tau \geq
10^4$ for $\gamma=10^{-3}$, and $\tau \geq 10^6$ for
$\gamma=10^{-5}$. We call this $\tau$-independence to be saturation
of the profile $J(x,\tau)$ around resonant frequency. In saturated
state, the number of photons redshifted from $\nu>\nu_0$ to the local Boltzmann
distribution area $\simeq \nu_0$ due to Hubble expansion is equal
to the number of photons leaving from $\nu_0$ to the red wing. Therefore, we see from
Figure 5 that once $J$ reaches the saturation state, the boundary
on the red wing $(x<0)$ of $J$ is moving to left (red). On the other
hand, the boundary on the blue wing $(x>0)$ is almost
time-independent.

Unlike in Figure 2, the width $x_l$ does not always increase with
time. For $\gamma=10^{-3}$ the width stops to increase when $\tau > 10^3$,
and for $\gamma=10^{-5}$, it is stopped at $\tau > 10^5$. One can 
find the mean scattering number $N_c$ with the similar way as
eq.(\ref{eq15}). When $\gamma \neq 0$, the time duration of photons
staying in the frequency space from $x$ to $x+\Delta x$ roughly is
$\Delta x/\gamma$ [eq.(\ref{eq11})]. On the other hand, the mean probability 
of $|x|\leq x_l$ photons being scattered in a unit $\tau$ is
$\displaystyle\frac{1}{x_l}\int_{0}^{x_l}\phi(x)dx$. The larger the $x_l$, 
the less the probability. Thus, all photons within $|x|<x_l$ averagely 
undergo $N_c$ scattering give by 
\begin{equation}
\label{eq16}
 N_c \simeq \frac{1}{\gamma}\int_{0}^{x_l}\phi(x)dx.
\end{equation}
 From Figure 5, the maximum width for
$\gamma=10^{-3}$ is estimated as $x_l=1.9$, corresponding to
$N_c \simeq 0.5\times 10^3$. While for $\gamma=10^{-5}$, maximum
width is $x_l=2.8$, and $N_c \simeq 5.0\times 10^5$. Once the width $x_l$ stops to
increse, all quantities in eq.(\ref{eq16}), $\gamma$, $x_l$ and $\phi(x)$, 
are $\tau$-independent. Thus, $N_c$ should also be $\tau$-independent.

\begin{figure}[htb]
\begin{center}
\includegraphics[width=6.5cm]{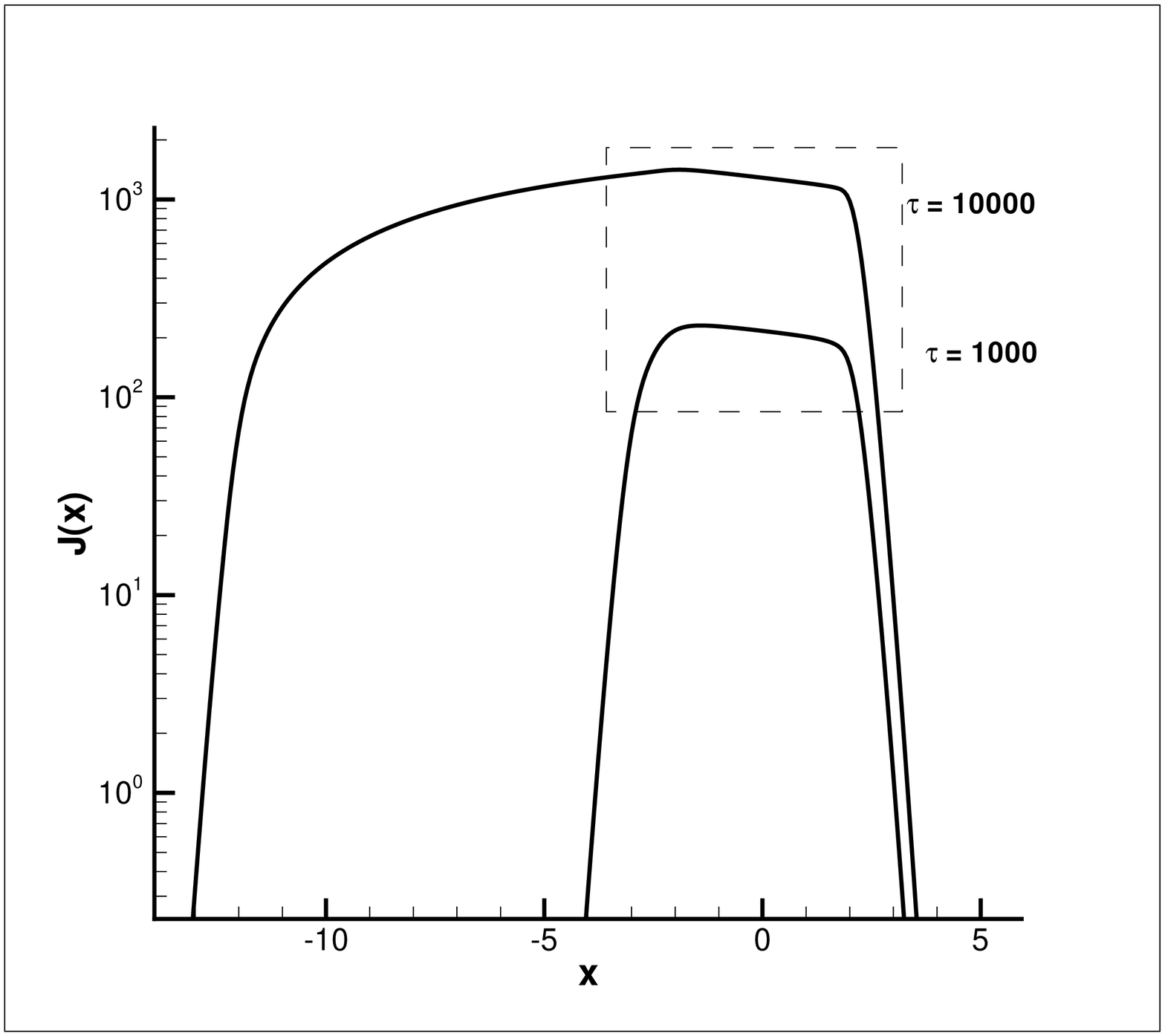}
\includegraphics[width=6.5cm]{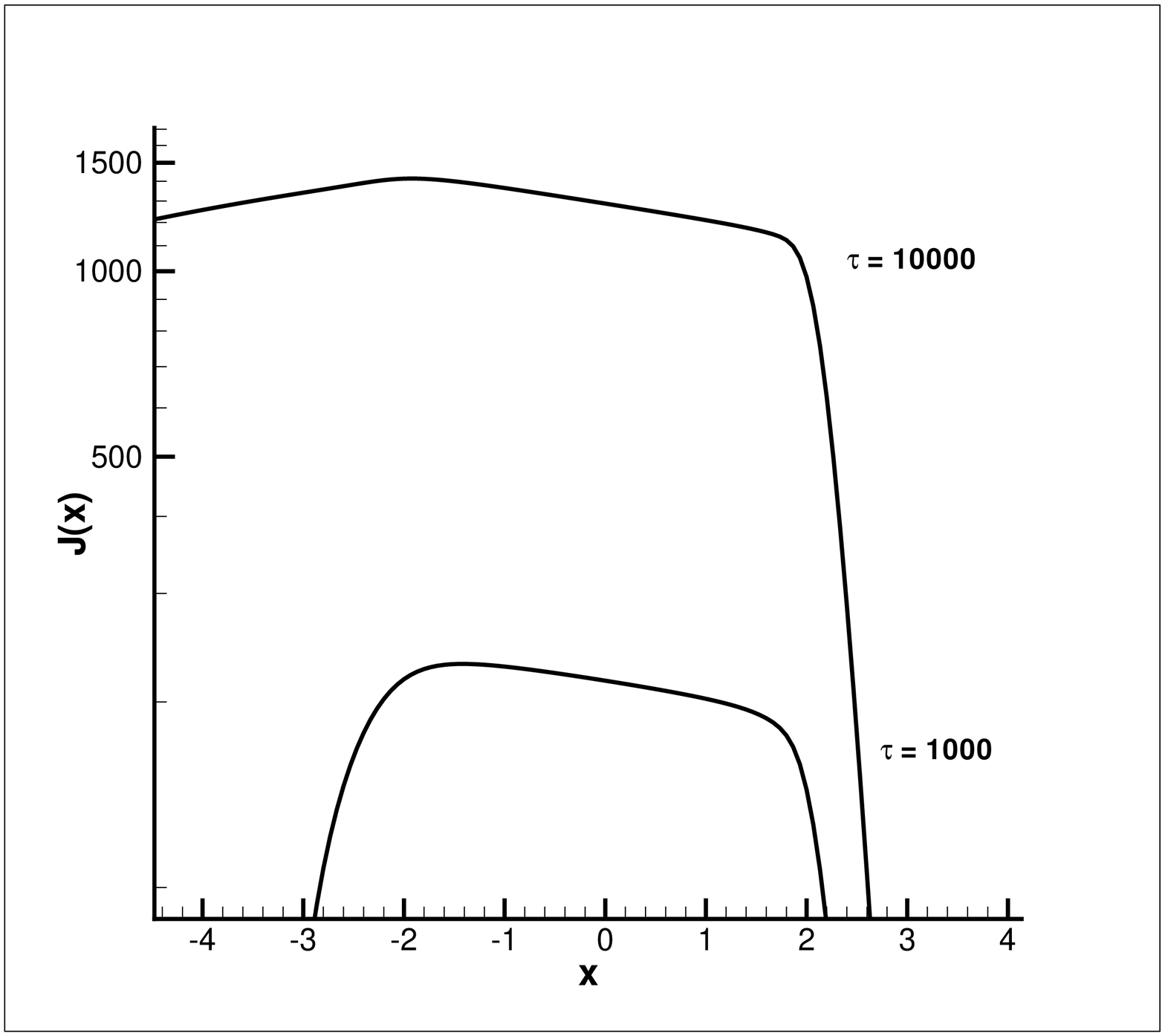}
\end{center}
\begin{center}
\includegraphics[width=6.5cm]{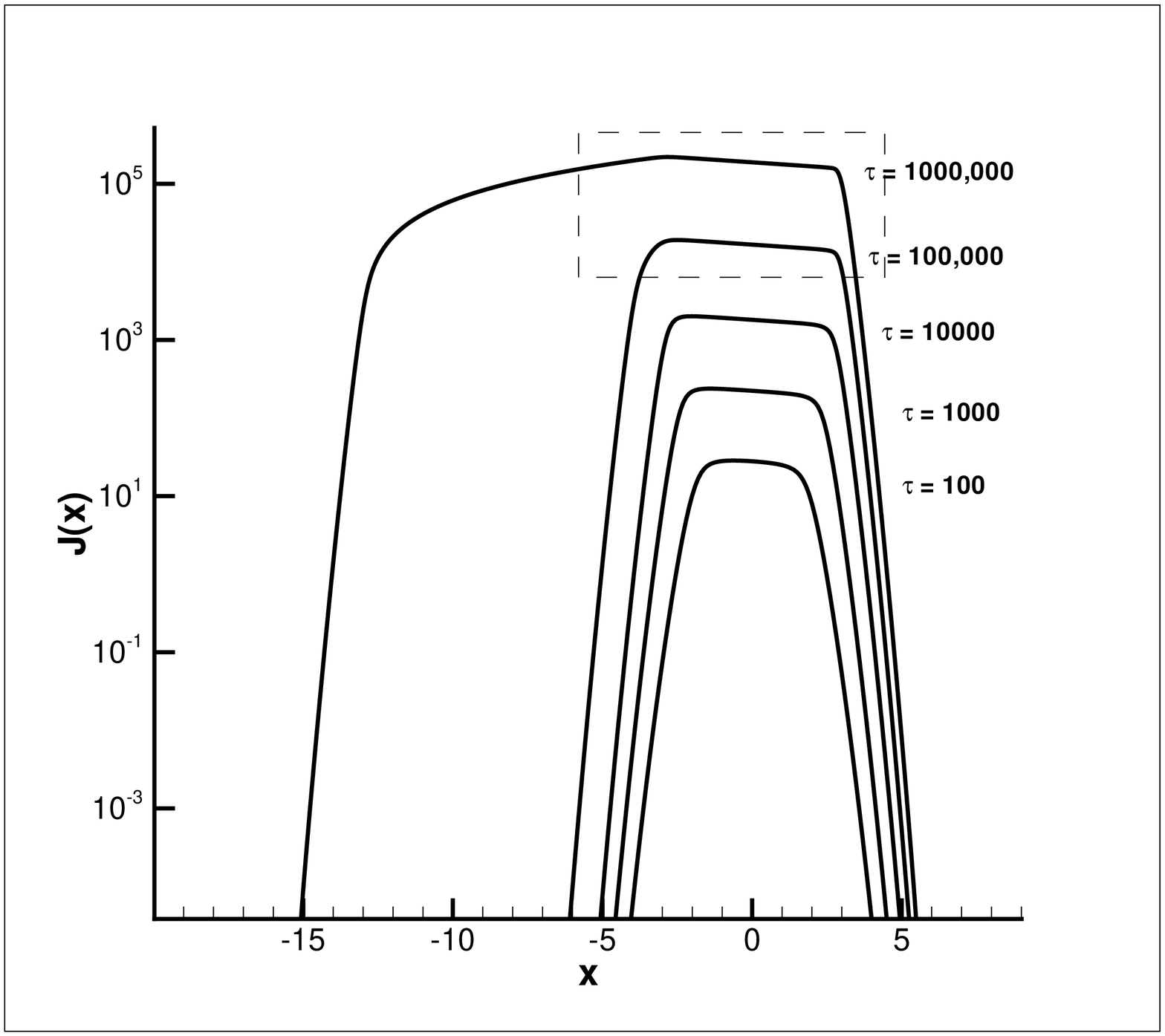}
\includegraphics[width=6.5cm]{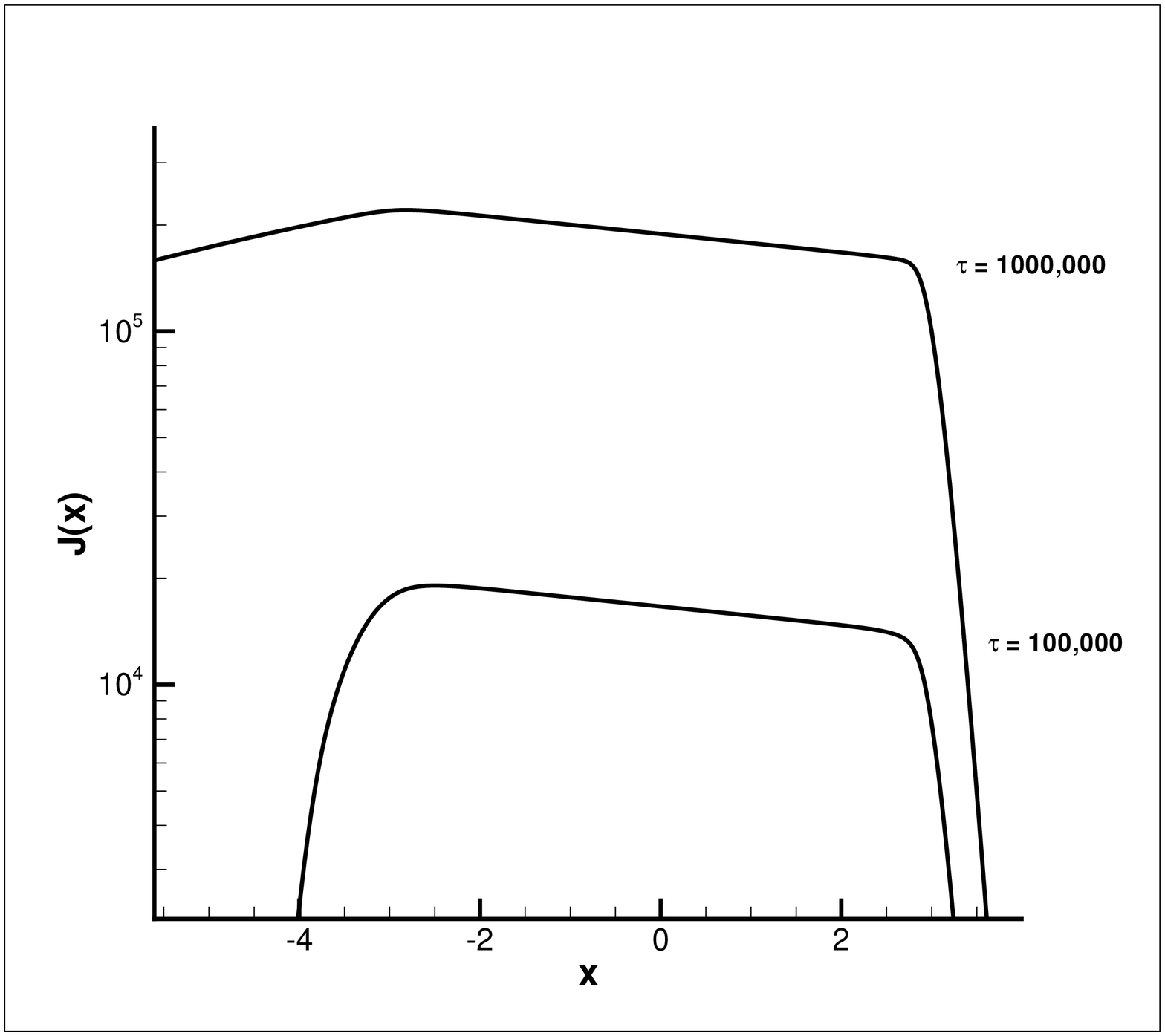}
\end{center}
\caption{WENO numerical solutions of eq.(\ref{eq9}) with $b=0.03$
and $\gamma=10^{-3}$ (two top panels) and $\gamma=10^{-5}$ (two
bottom panels). The source is taken to be $S=\phi(x)$ and initial
condition $J(x,0)=0$. The right panels are zoom-in of the dashed
square of the corresponded left panel.} \label{fig5}
\end{figure}

The $\tau$-independence of the width $x_l$ is also shown in Figure
6, in which we still use $\gamma=10^{-3}$, and $J(x,0)=0$ initially.
However, the source is taken to be $S(x)=\phi(x-10)$. That is, the
source photons have frequency $x=10$, or $\nu=\nu_0+10\Delta \nu_D$.
The resonant scattering at $x=0$ ($\nu=\nu_0$) will occur when these
photons have redshifted from $\nu$ to $\nu=\nu_0$, which takes time
of about $\tau=10/\gamma\simeq 10^4$. Figure 6 shows that the whole
distribution of $J(x,\tau)$ dramatically evolves with time, but the
width of the local Boltzmann distribution around $x=0$ is kept to be
$x_l\simeq 2$ from $\tau=10^4$ to $4\times 10^4$. We also
find from our numerical calculations that when $\tau>4\times
10^4$, the intensity of the photon flux around $x=0$ keeps constant,
or it is in saturated state.
\begin{figure}[htb]
\begin{center}
\includegraphics[width=4.5cm]{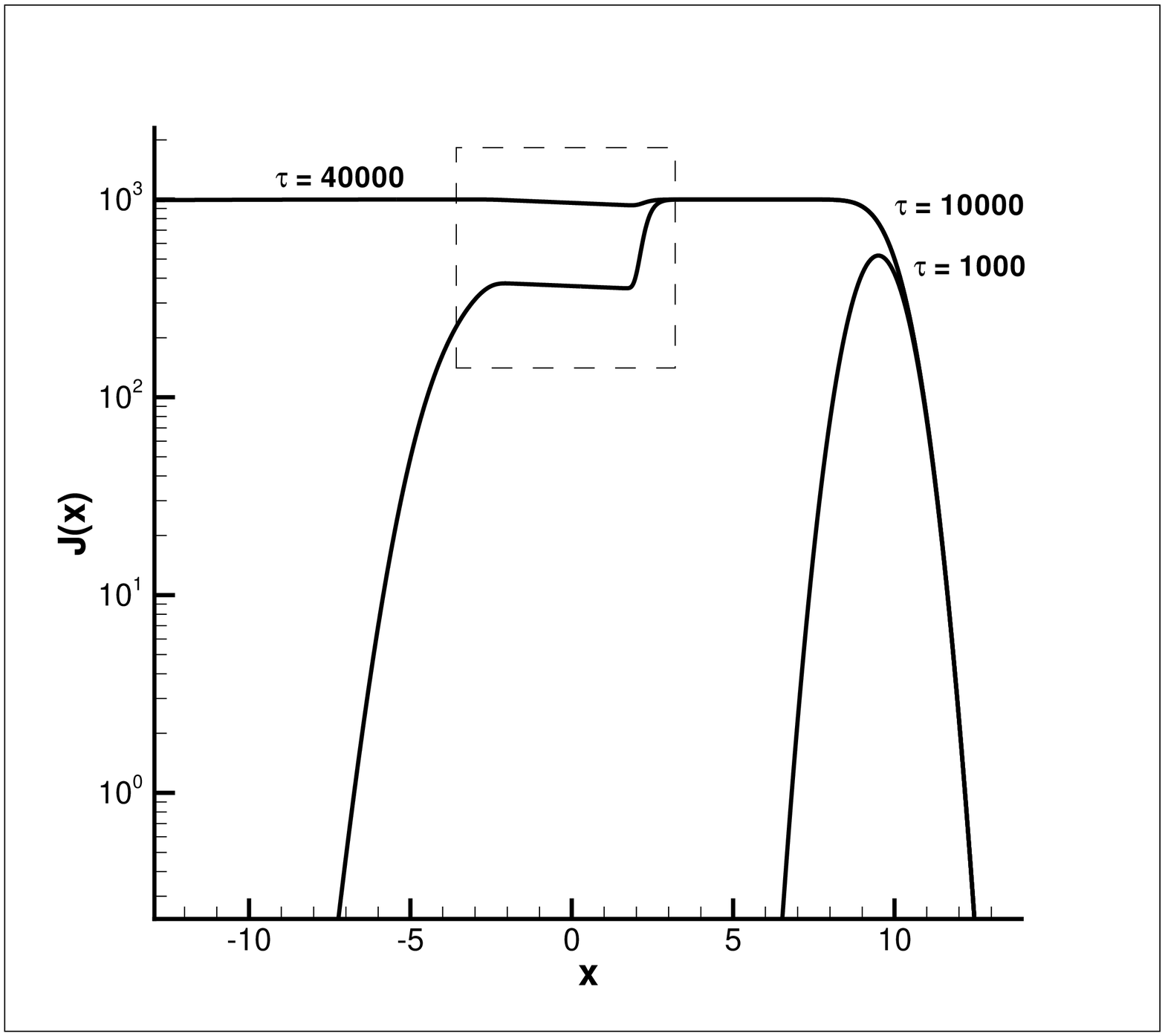}
\includegraphics[width=4.5cm]{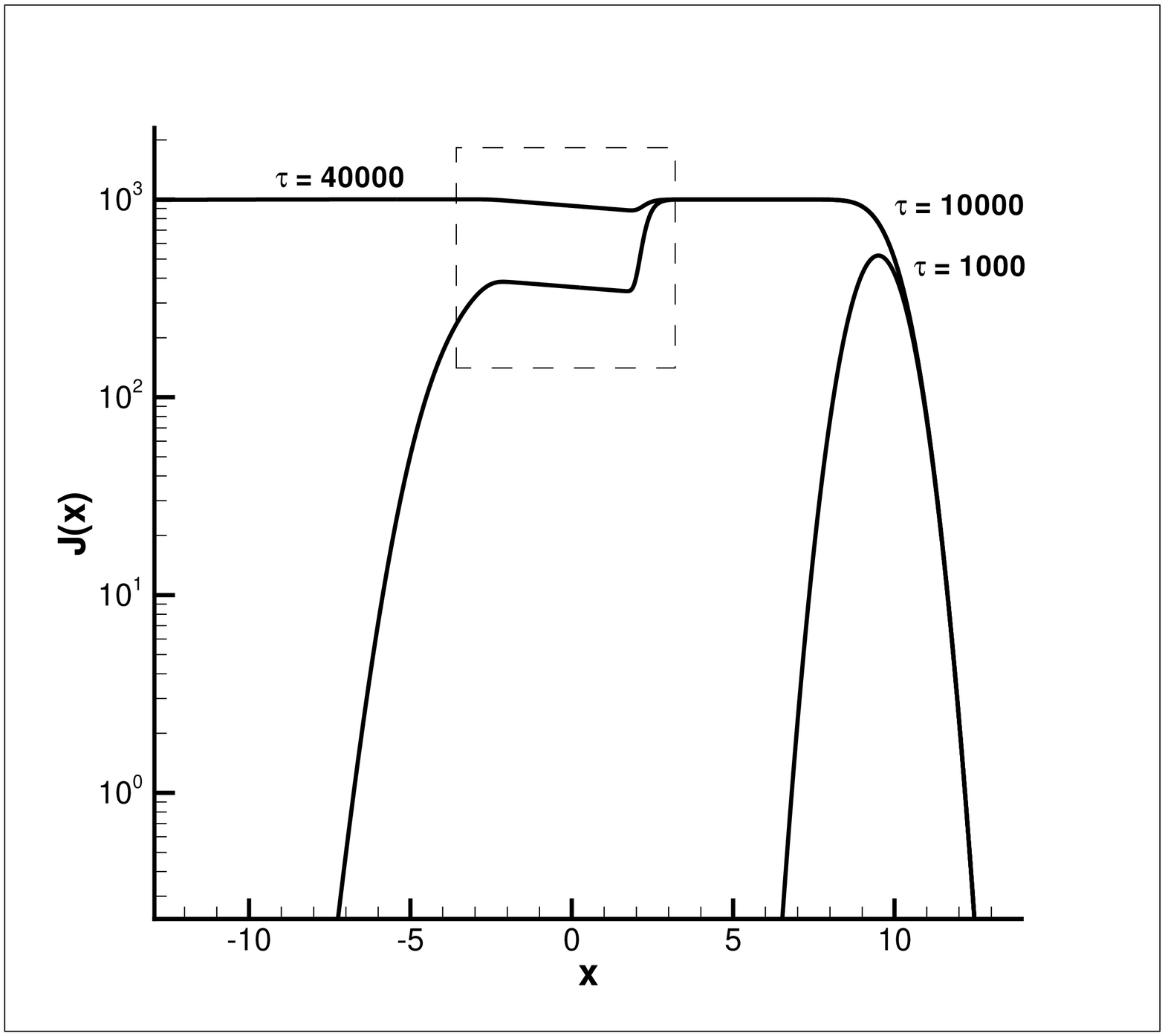}
\includegraphics[width=4.5cm]{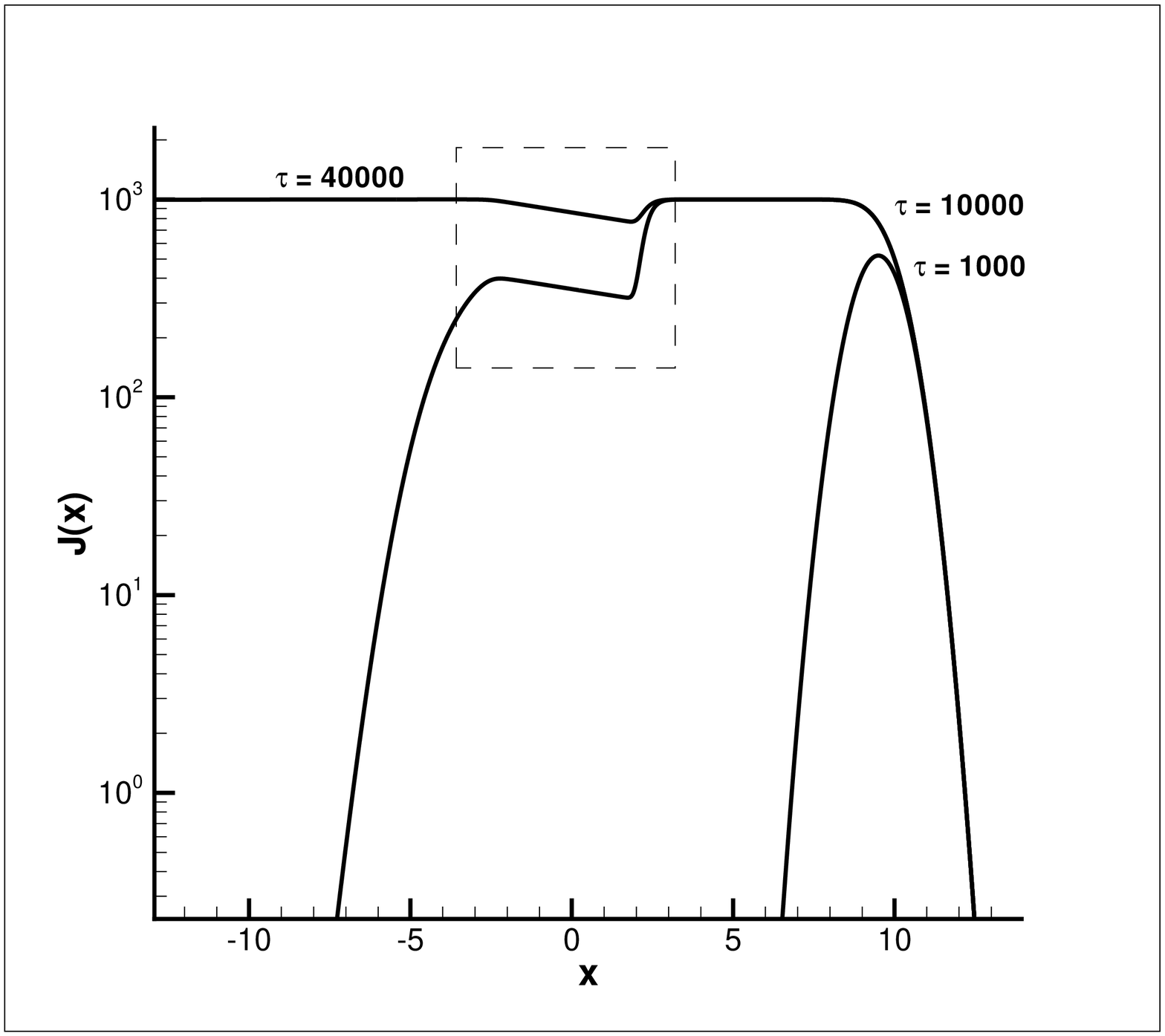}
\end{center}
\begin{center}
\includegraphics[width=4.5cm]{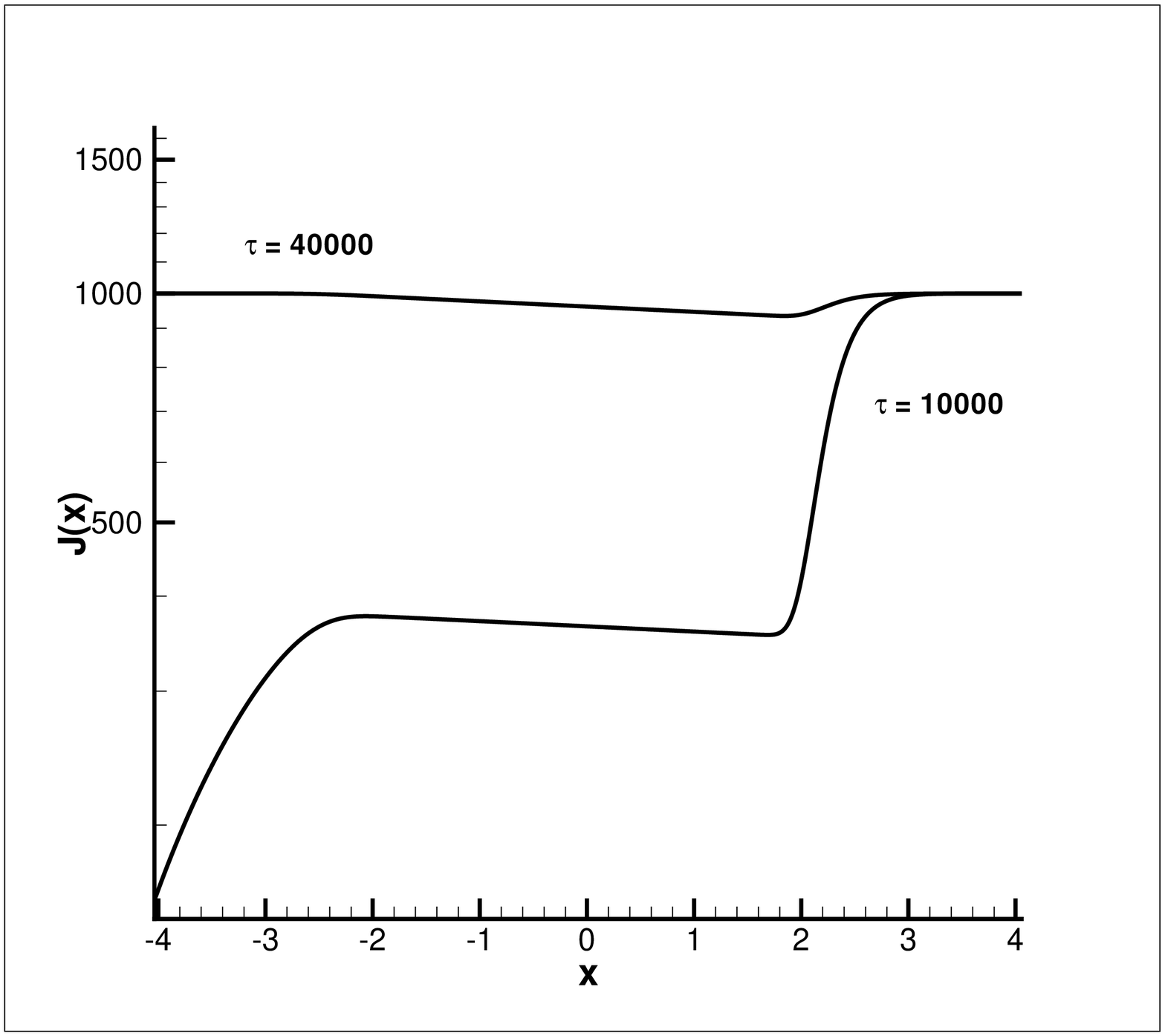}
\includegraphics[width=4.5cm]{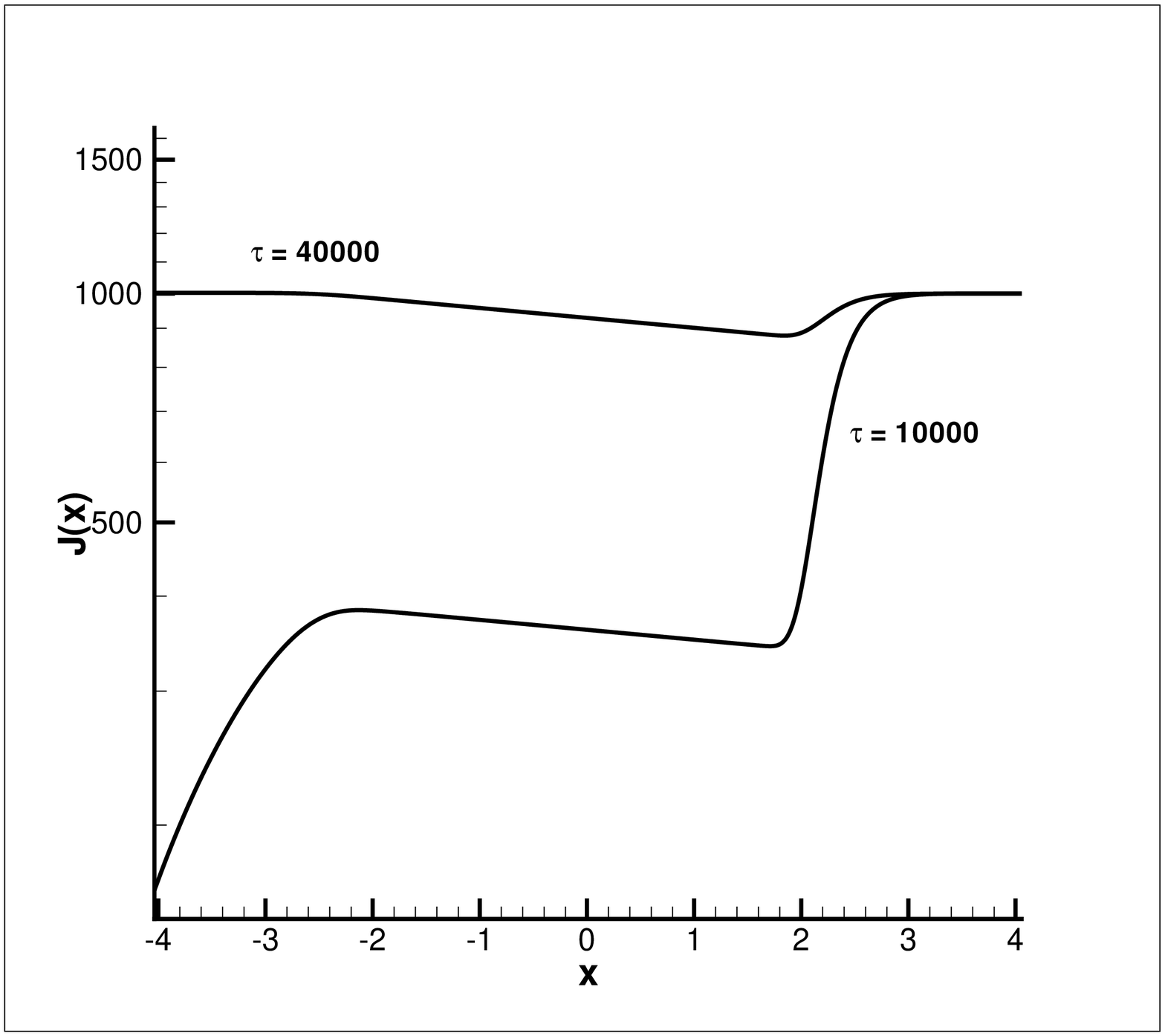}
\includegraphics[width=4.5cm]{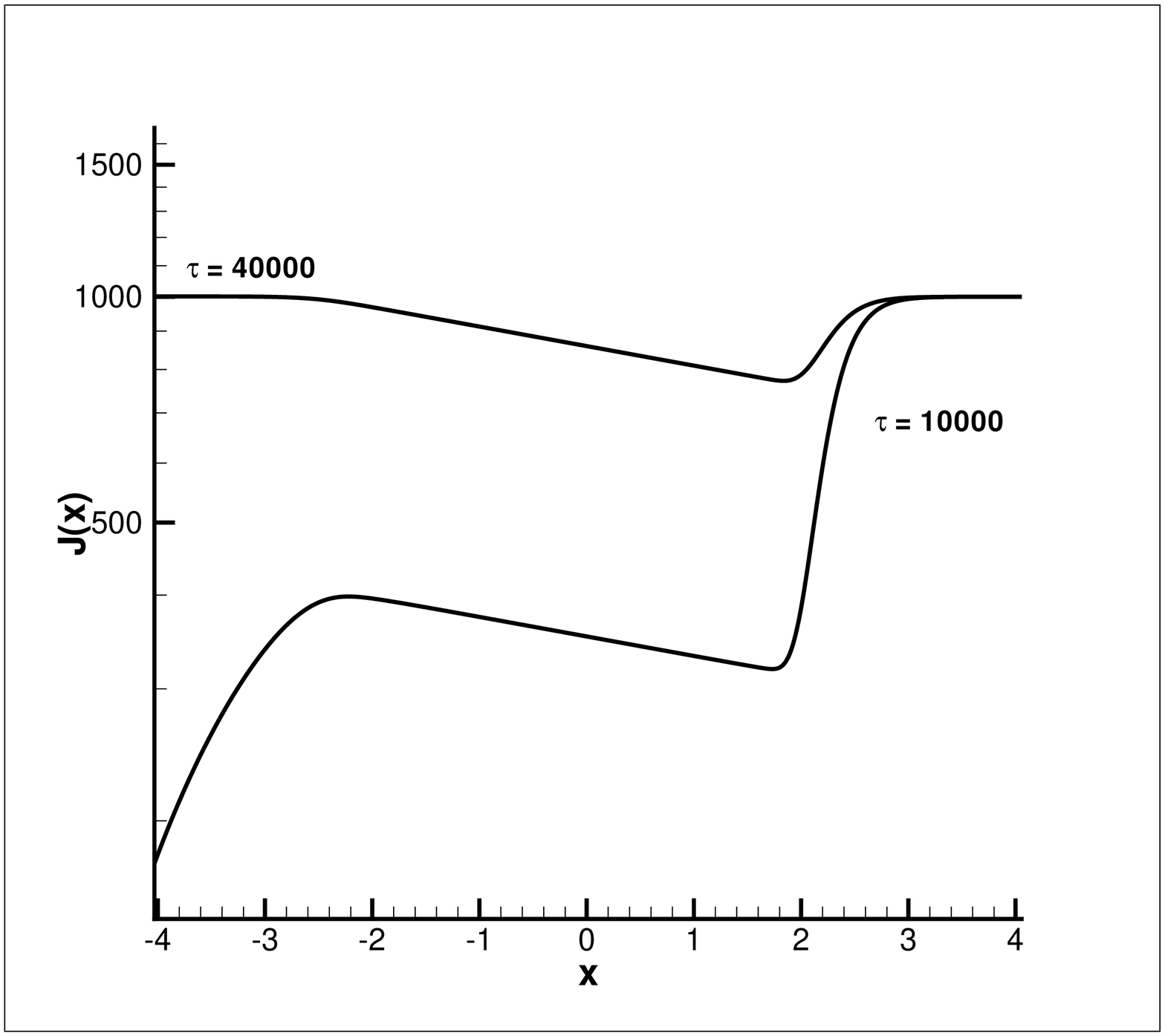}
\end{center}
\caption{WENO numerical solutions of eq.(\ref{eq9}) with $\gamma=10^{-3}$
and $b=0.0079$ (left), 0.015 (middle), 0.03 (right). The source is
taken to be $S=\phi(x-10)$ and initial condition $J(x,0)=0$. The
bottom panels are the zoom-in of the dashed square of the top panels. }
\label{fig6}
\end{figure}

Similar to Figure 3, Figure 6 also shows that the width of the local
Boltzmann distribution is approximately $b$-independent. From
eq.(\ref{eq16}), one can also expect that the width will be smaller
for larger $\gamma$. A local Boltzmann distribution can form only if
$\gamma^{-1}$ is large enough. This property is shown with Figure 7,
in which we use the same photon source $S(x)$ and parameter $b$ as
in Figure 6, but we take larger $\gamma$. Figure 7 presents the
results of $\gamma=1$ and $10^{-1}$. We see from Figure 7 that in
the case of $\gamma=1$, there is no local Boltzmann distribution at
any time. The resonant scattering leads only to a valley around
$x=0$. It is because the strongest scattering is at $x\simeq 0$, which
moves photons with frequency $x\simeq 0$ to
$x\neq 0$. However, the redshift lets photon quickly leaving from
$x\simeq 0$. They are not undergoing enough number of scattering to
form a local statistical equilibrium distribution. For
$\gamma=10^{-1}$, it seems to show a local Boltzmann distribution,
but its width is very small at all time.
\begin{figure}[htb]
\begin{center}
\includegraphics[width=5.0cm]{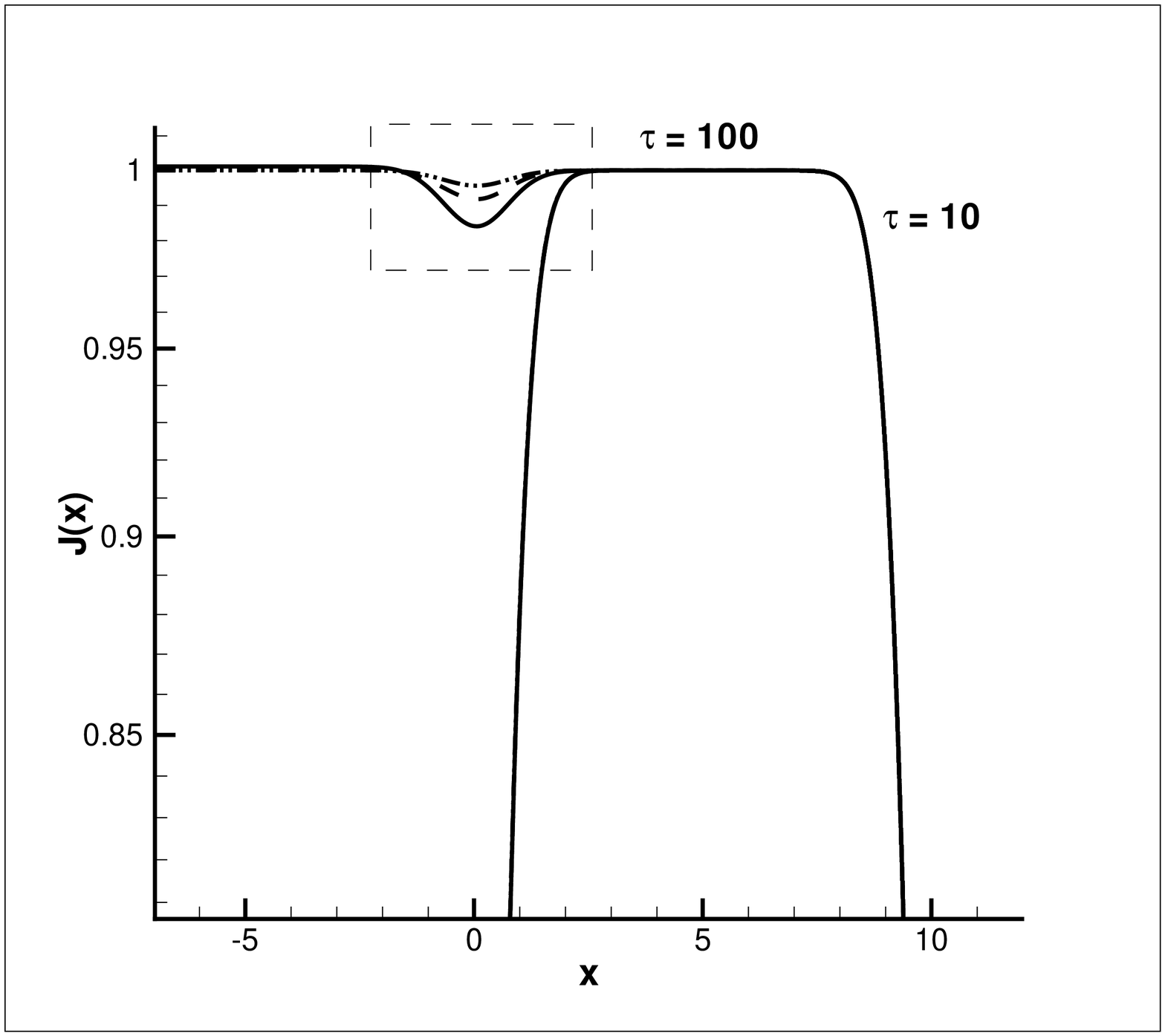}
\includegraphics[width=5.0cm]{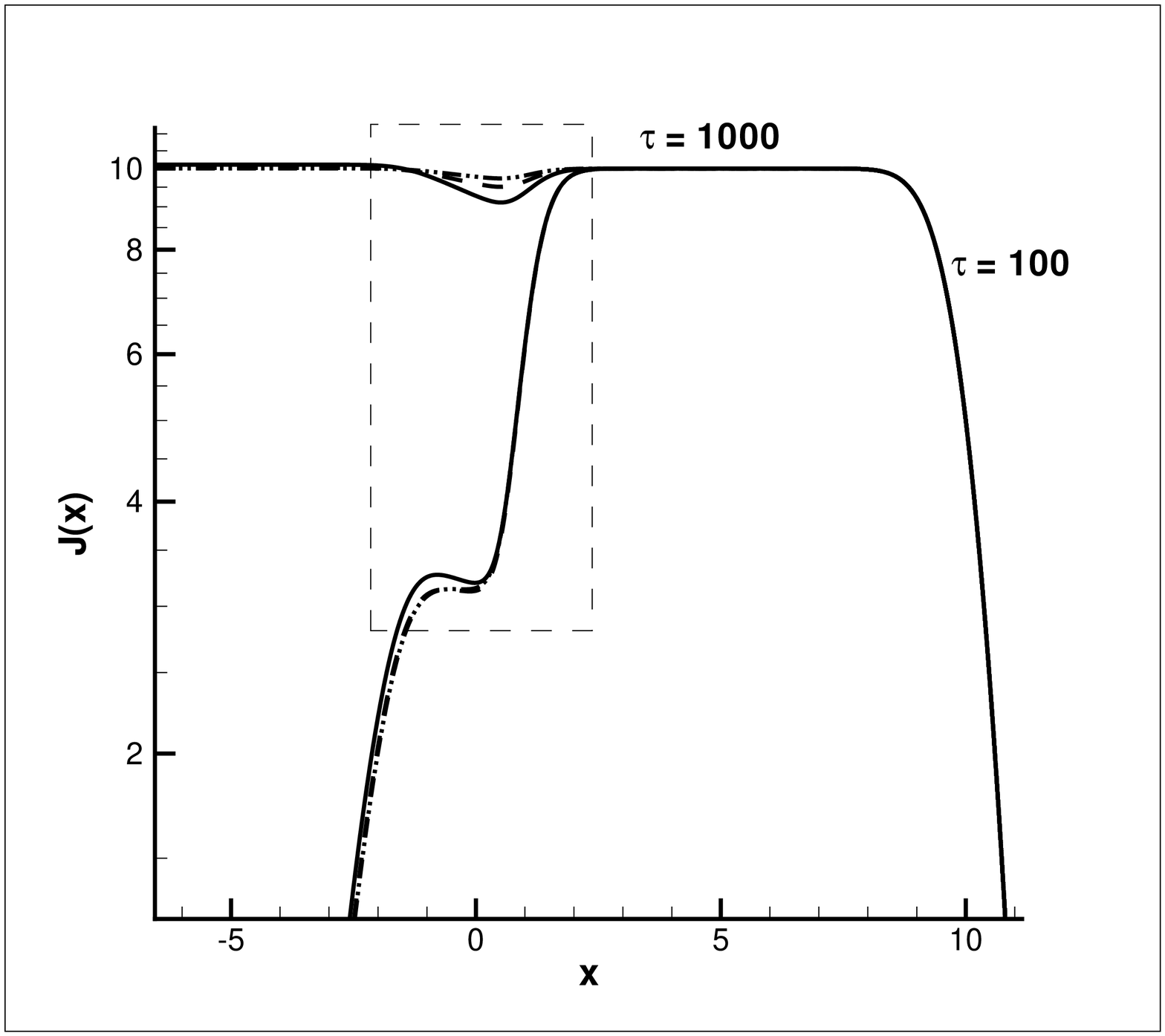}
\end{center}
\begin{center}
\includegraphics[width=5.0cm]{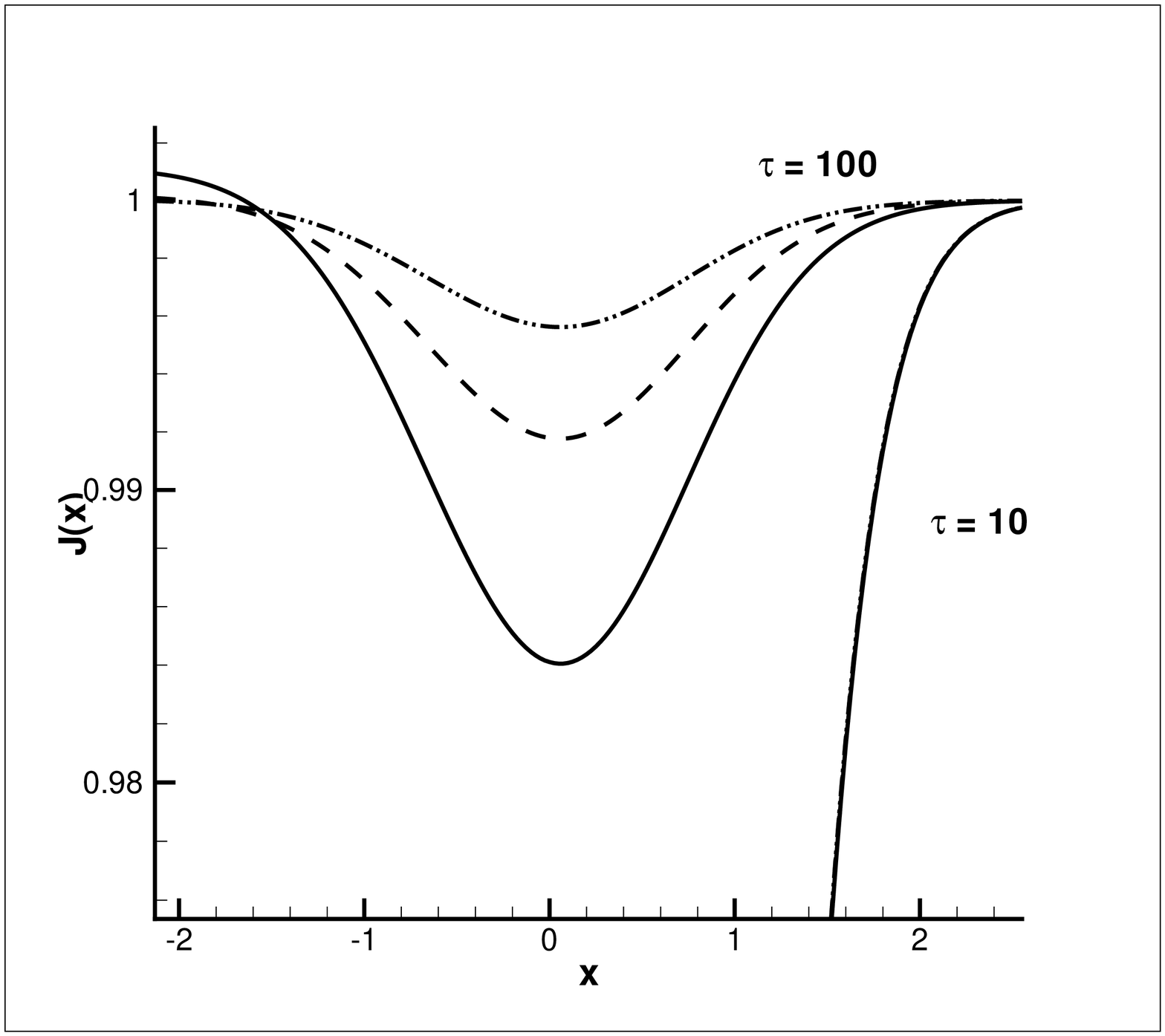}
\includegraphics[width=5.0cm]{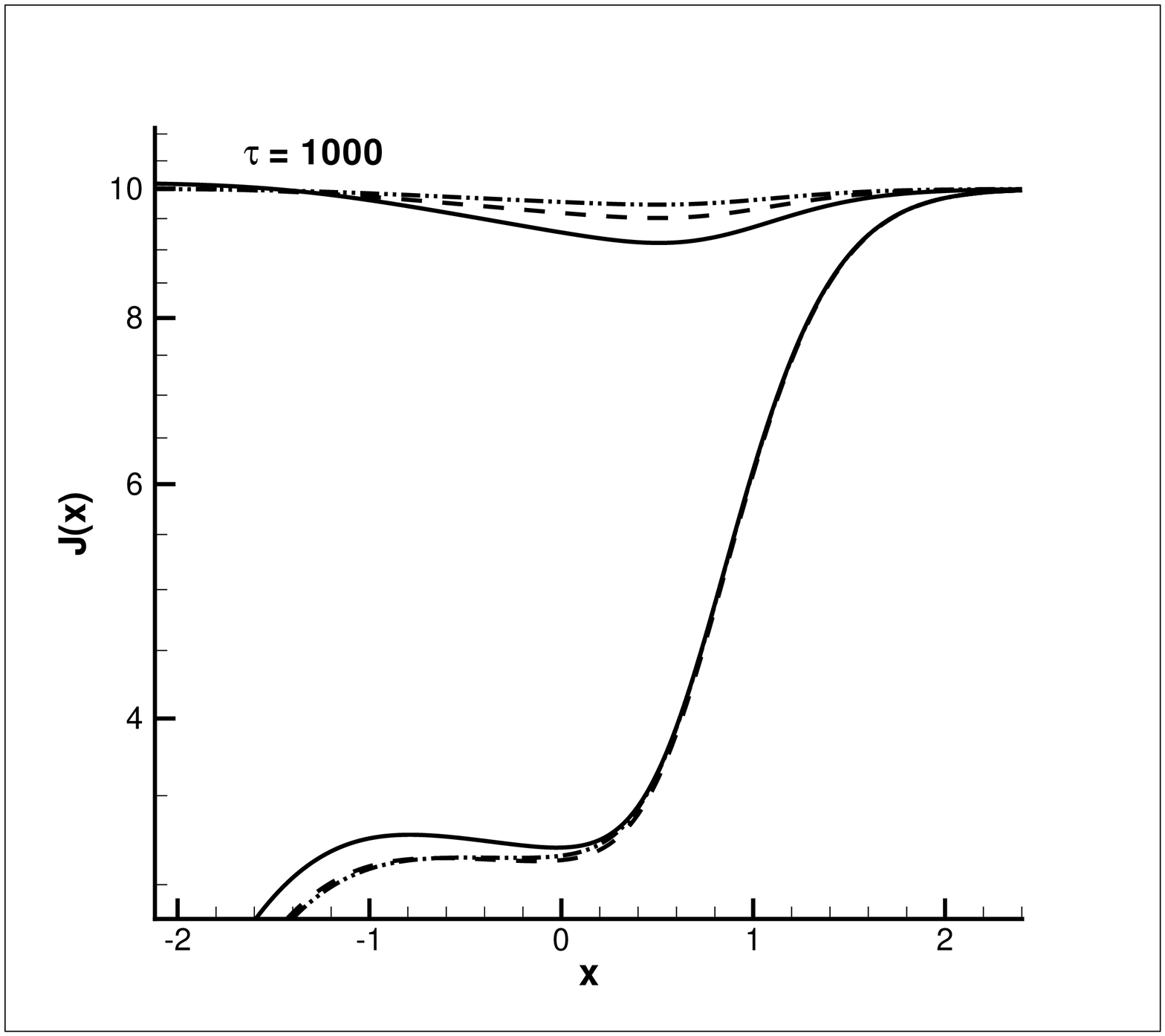}
\end{center}
\caption{WENO numerical solutions of eq.(\ref{eq9}) with $\gamma=1$ (left)
and $\gamma=10^{-1}$ (right). Parameter $b=0.0079$ (dot-dot-dashed), 0.015
(dashed), 0.03 (solid). The
source is $S=\phi(x-10)$ and initial condition $J(x,0)=0$. The
bottom panels are the zoom-in of the dashed square of the top panels.}
\label{fig7}
\end{figure}

\subsection{Photon source and W-F coupling}

The $\tau$- and $b$-independencies of the shape and width of the
local Boltzmann distribution yield an important conclusion that for
given parameters $\gamma$ and $b$, the formation and evolution of
the local Boltzmann distribution is independent of the photon
sources $S(x, \tau)$. This is because eq.(\ref{eq9}) is linear of
$J$. Any source $S(x, \tau)$ can always be considered as a
superposition of many monochromatic sources around frequency
$x=x_i$, or $S(x,\tau)=\sum_{i}S_i\phi(x-x_i,\tau)$, $S_i$ is the
intensity of photon source $\phi(x-x_i,\tau)$ with frequency $x_i$.
$J(x,\tau)$ can be decomposed into $J(x,\tau)=\sum_{i}J_i(x,\tau)$,
where $J_i(x,\tau)$ is the solution of eq.(\ref{eq9}) with the
source $i$. Thus, if the formation of the local Boltzmann
distribution around $x=0$ is independent of $\tau$ and $b$, the
superposition $J(x,\tau)=\sum_{i}J_i(x,\tau)$ should also show the
same local Boltzmann distribution around $x=0$. Although the overall
amplitude does depend on the source, the shape around resonant frequency
does not.

As an example, Figure 8 presents a solution with the same parameters
as in Figure 6, but the source is with continuous spectrum given by
\begin{equation}
\label{eq17}
S(x,\tau) =\left \{ \begin{array}{ll}
         (10/x)^{2}, & 10 < x < 15, \\
         0 & {\rm otherwise}  \end{array}
         \right .
\end{equation}
Photons with frequency $x=10$ will arrive earlier at $x=0$ with
higher intensity, while those with frequency $x=15$ will arrive at
$x=0$ later with lower intensity. The flux $J(x,\tau)$ of Figure 8
has very different shape from Figure 6, while the local Boltzmann
distribution at $-2<x<2$ of Figure 8 is exactly the same as that in
Figure 6. Therefore, the W-F coupling is always working regardless
the original spectrum of the redshifted photons.
\begin{figure}[htb]
\centering
\includegraphics[width=5.0cm]{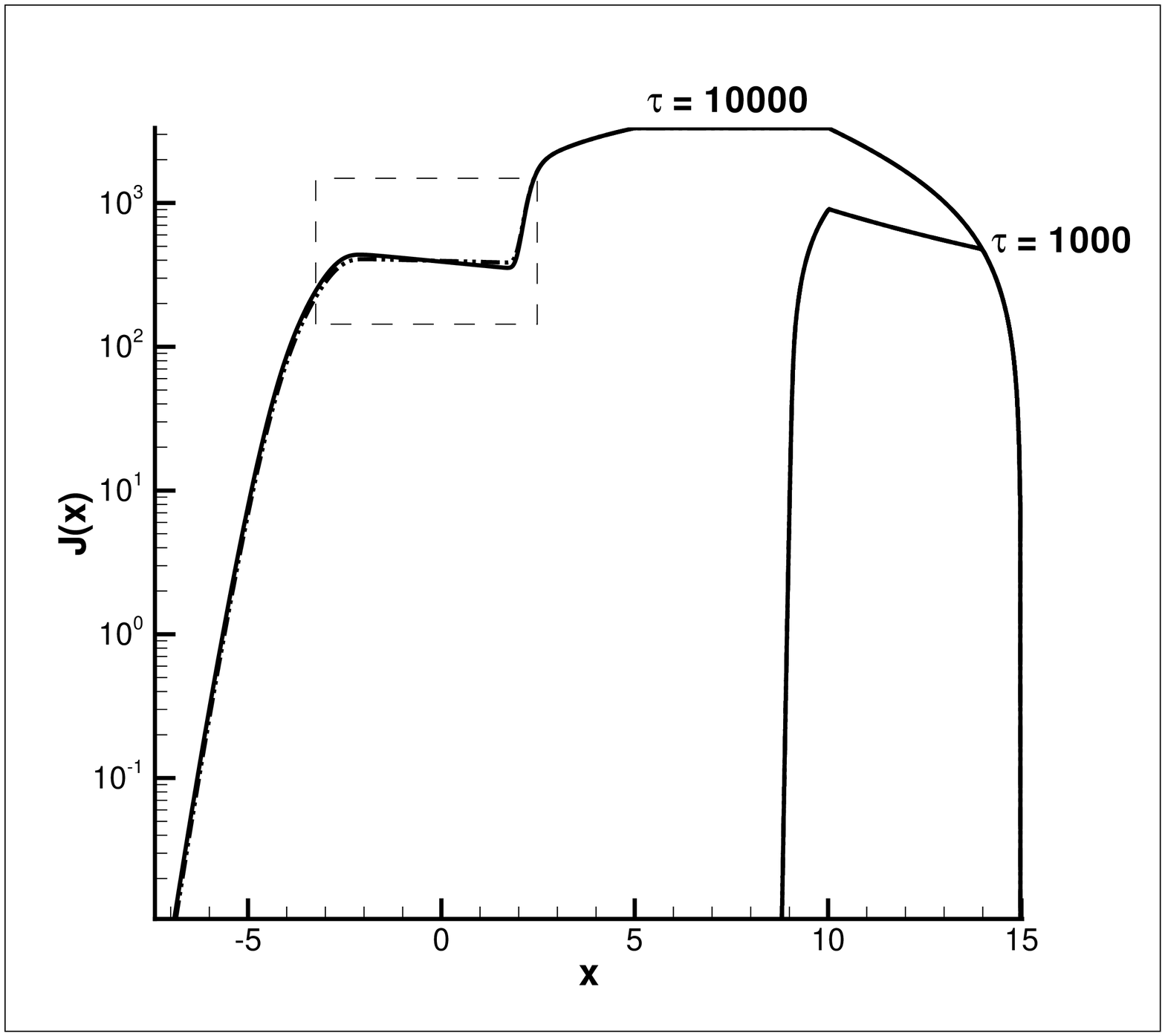}
\includegraphics[width=5.0cm]{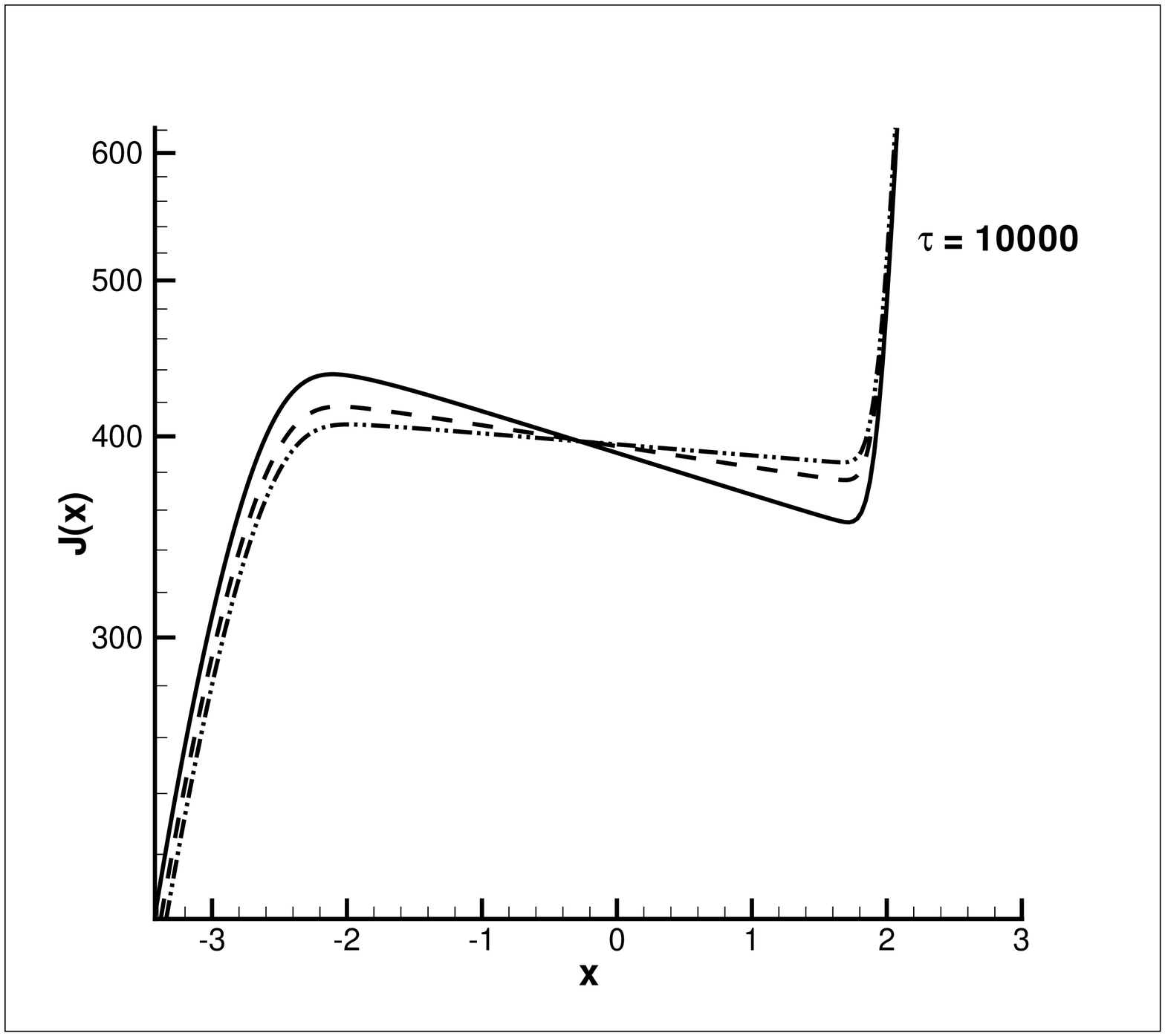}
\caption{WENO numerical solutions of eq.(\ref{eq9}) with $\gamma=10^{-3}$
and $b=0.0079$ (dot-dot-dashed), 0.015 (dashed), 0.03 (solid). The source
is given by eq.(\ref{eq17}). The
right panel is a zoom-in of the dashed square of left panel.}
\label{fig8}
\end{figure}

\subsection{Intensity}

 From Figures in \S 5.1 and 5.2, we see that the intensity of
photon flux $J$ at the local Boltzmann distribution is strongly
dependent on $\gamma$ and $\tau$. Figures 5 and 6 show that at early
time $J$ is smaller than its saturation state. For the solution of
Figure 6, the flux in the frequency range of the local Boltzmann
distribution is saturated at about $\tau = 4\times 10^4$ with
saturated flux $J \simeq 10^4$, while the intensity at $\tau=10^4$
is significantly lower than that of the saturated state.

 Figure 9 is the same as Figure 6, but taking $\gamma=10^{-2}$ and
$10^{-4}$. The lower panels of Figure 9 shows once again that the
time-evolution of the intensity is about $b$-independent. Figure 9
shows also that for $\gamma=10^{-2}$, the photon flux approaches the
saturated state with intensity of $J \simeq 10^2$ at $\tau=10^4$.
While for $\gamma=10^{-4}$, the saturated state has not yet been
approached even when intensity $J\simeq 10^4$, and $\tau=10^5$.
Generally, the smaller the $\gamma$, the larger the saturated
intensity and the longer the $\tau$ needed to approach its
saturated state.

\begin{figure}[htb]
\centering
\begin{center}
\includegraphics[width=5.0cm]{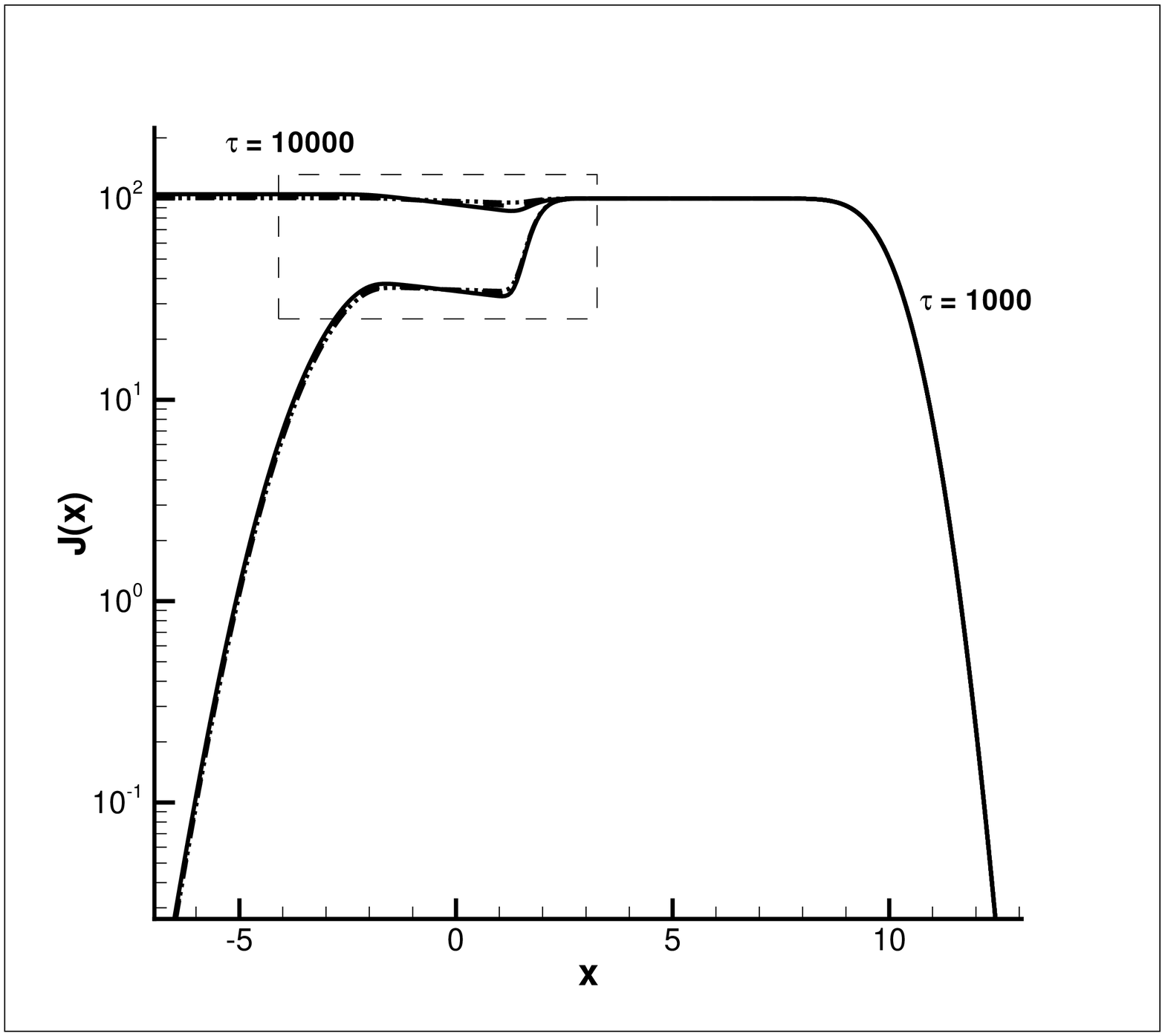}
\includegraphics[width=5.0cm]{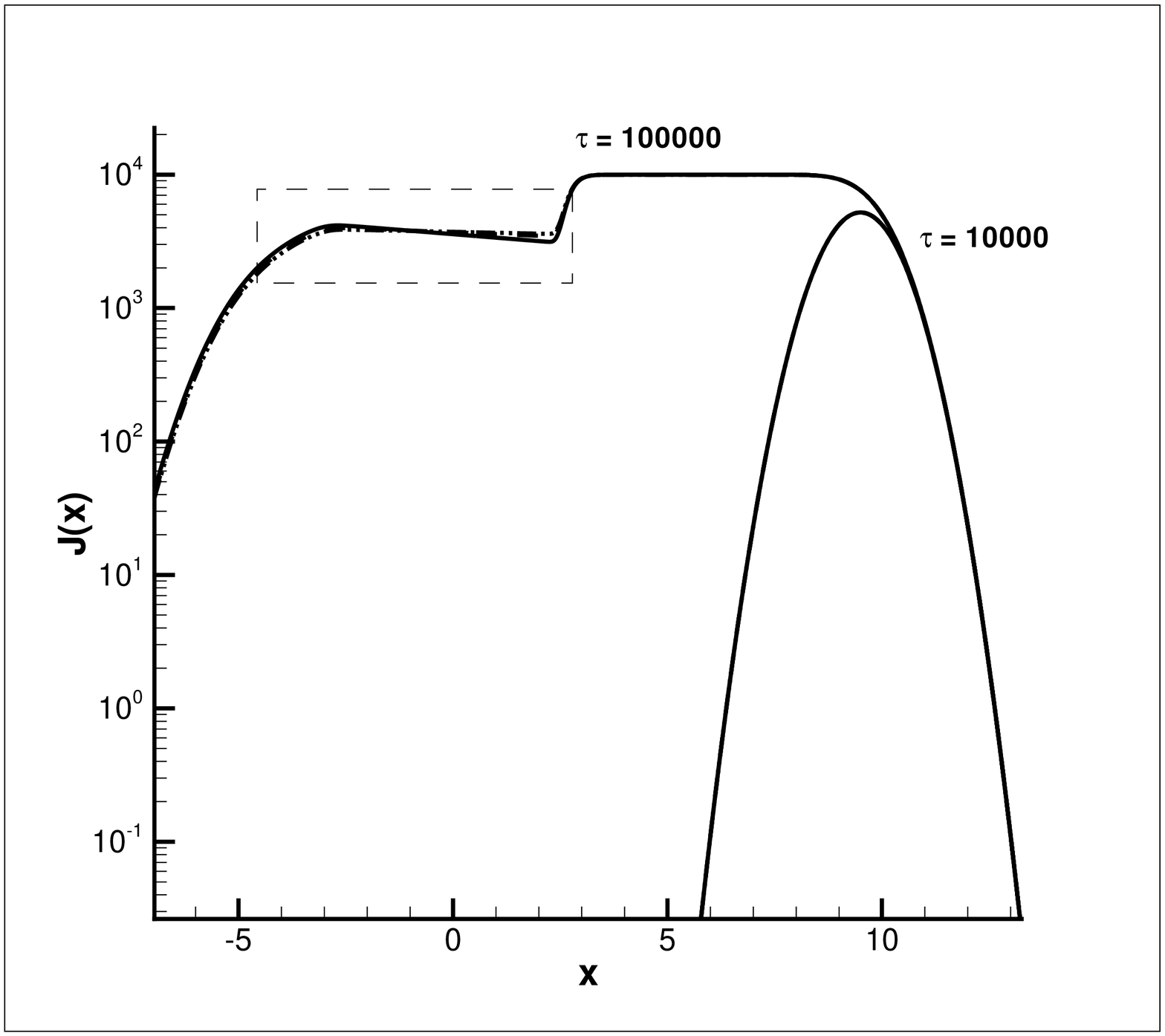}
\end{center}
\begin{center}
\includegraphics[width=5.0cm]{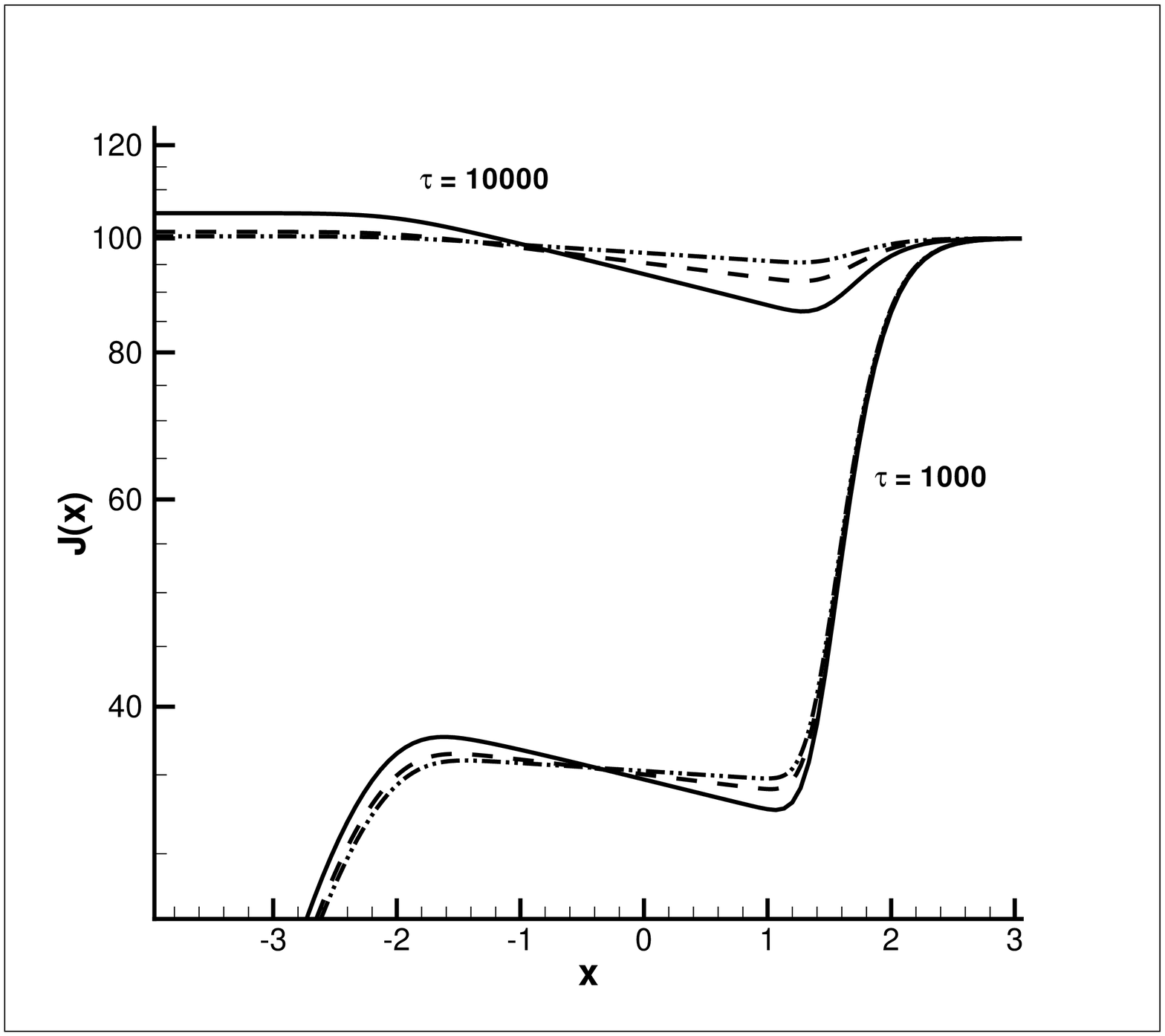}
\includegraphics[width=5.0cm]{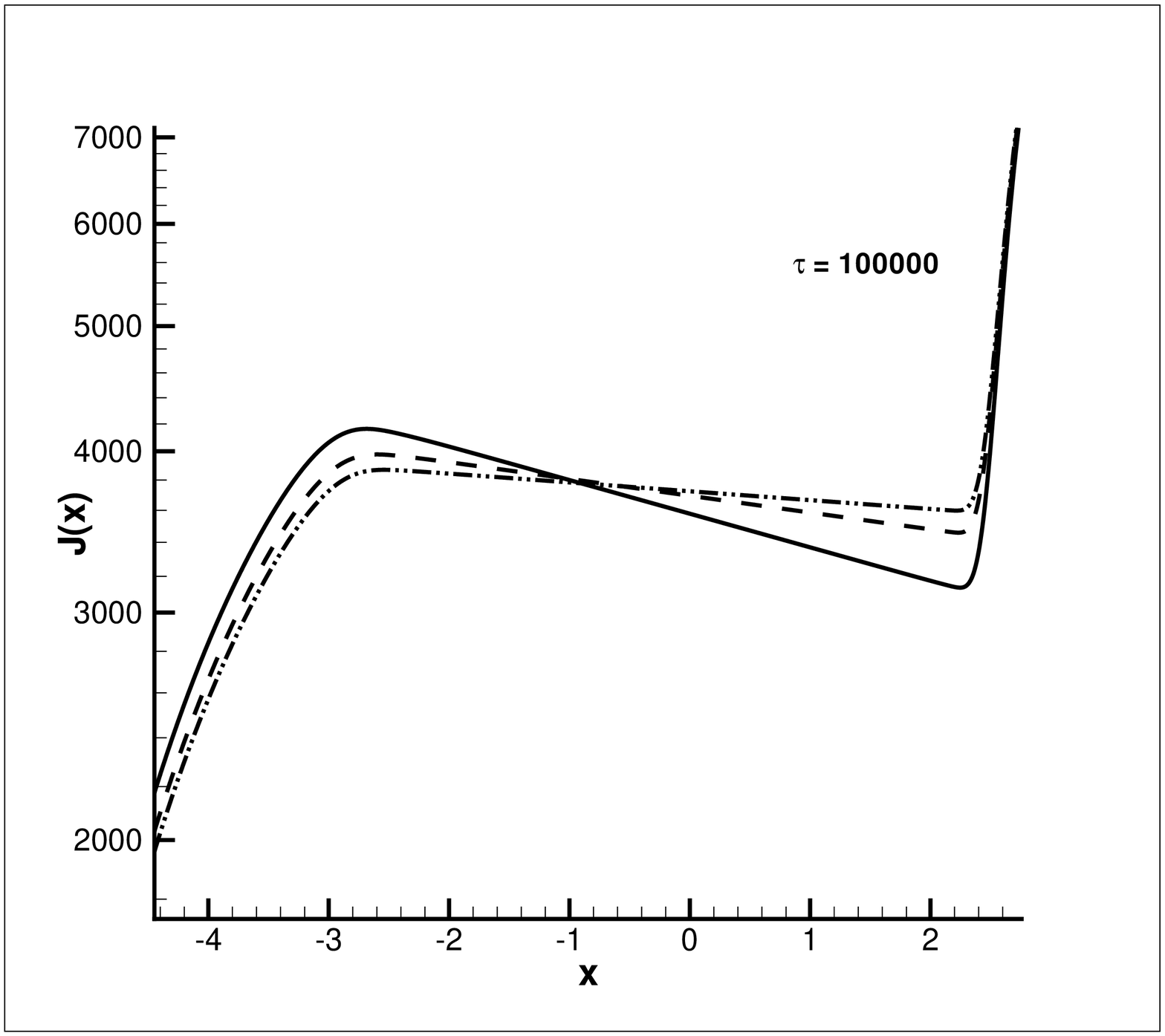}
\end{center}
\caption{WENO numerical solutions of eq.(\ref{eq9}) with $\gamma=10^{-2}$
(left) and 10$^{-4}$ (right) with source $S(x)=\phi(x-10)$ and
$b=0.0079$ (dot-dot-dashed), 0.015 (dashed), 0.03 (solid).  The bottom
panels are the zoom in of the
dashed square of top panels.} \label{fig9}
\end{figure}

\section{Conclusions and discussions}

\subsection{Summary}

The onset of the W-F coupling, or the formation of local Boltzmann
distribution is similar to the process of approaching a
statistically thermal equilibrium state via collisions or
scattering. The particle distribution in the statistical equilibrium
is independent of time, initial distribution, and the details of
collision. The equilibrium distribution is maintained only by the
enough collisions among particles. A local Boltzmann distribution is
formed once the number of resonant scattering is large enough. Like
other statistically thermal equilibrium, the features of the local
Boltzmann distribution are independent of time, photon source,
initial photon distribution, and etc.

In an expanding universe, photons are moving in the frequency space with
``speed" given by the redshift. The formation of the local Boltzmann
distribution depends on the competition between the resonant
scattering and the redshift. If photons have undergone enough scattering
during their path through the frequency space around the resonant
frequency, a local Boltzmann distribution will be formed. Otherwise
local statistical equilibrium cannot be approached.

In our work, we use the Gaussian profile eq.(\ref{eq3}), but not the
Voigt profile. Our major results on the formation of local Boltzmann
distribution will not be affected by using the Voigt profile if the
width $x_l$ is not larger than the Doppler thermal broadening. We
found that the solution of eq.(9) with rest background is not
affected by Voigt profile if the ratio $a$ between the natural and
Doppler broadening is equal to 10$^{-3}$, which corresponds to
$T\simeq 650$ K. For large $a$, or small $T$, the Doppler
thermal broadening is small. In this case, the width $x_l$, in which 
the color temperature $T_c$ of Ly$\alpha$ photons is locked to the 
kinetic temperature $T$ of hydrogen atoms, would also be small.

\subsection{Applications to the 21 cm problem}

A basic problem for the 21 cm signal from the ionized and heated
region around first stars is the conditions on which one can
estimate the 21 cm emission and absorption with the W-F coupling,
which forces the internal (spin-)degree of freedom to be determined
by the thermal motion of the atoms. In this case, the relative
occupation of the two hyperfine-structure components of the ground
state depends only upon the shape of the spectrum near the
Ly$\alpha$ frequency. Therefore, we need a local Boltzmann
distribution of Ly$\alpha$ photon with frequency width equal to or
larger than $|\nu_f -\nu_0|\geq \nu_{21}=1420$ MHz, or
\begin{equation}
x_f \geq 0.014 \left (\frac{10^4}{T}\right)^{1/2}.
\end{equation}
Thus, from Figure 7 and eq.(10) one can conclude that in regions
with $f_{\rm HI}\leq 10^{-4}$, where $\gamma > 10^{-1}$, the W-F coupling
will not work. Although electron-hydrogen and proton-hydrogen collisions
can be important; the 21 cm signal will be incredibly small. On the other hand,
in the primarily neutral IGM, W-F coupling is very efficient.

 From eqs.(7) and (10), we have
\begin{equation}
t=0.26\times 10^5 h^{-1} \left(\frac{T}{10^4}\right)^{1/2}\left (\frac{10}{1+z}\right
)^{3/2}\left(\frac{0.25}{\Omega_M}\right)^{1/2}\gamma \tau \hspace{3mm} yrs.
\end{equation}
We know that the saturation, or time independent solution around
resonant frequency, can be used only when the time $\tau$ is larger
than the time of photon moving over a frequency space from $-x_l$ to
$x_l$. From Figures 6 and 9, we have $\tau \gamma$ equal to about 10 at the saturation.
 Therefore, eq.(19) yields that the time independent solution
is available only if the time scale of the evolution of the ionized sphere
of first stars is larger than about $3\times 10^5$ yrs. This gives a constraint on
the 21 cm emission region, as such regions are very narrow, the time
scale of the evolution being comparable to $10^6$ yr, or even less
(Liu et al. 2007).

Actually we may not need a saturation state. What we used for
estimating the 21 cm signals is the frequency distribution
$J(x,\tau)$ to show a Boltzmann-like shape in the central part
$|x|\simeq 0$, i.e. the onset of the W-F coupling. The time-scale of
the W-F coupling onset is equal to about 10$^3$ yrs for neutral
hydrogen density of the concordance $\Lambda$CDM model [eq.(7)].
This time scale is much less than that of the evolution of 21 cm
region of first stars. It seems to indicate that we can safely use
the W-F coupling in the 21 cm estimation of first stars.

However, we should mention the effect of the photon intensity. The
coupling coefficient between Ly$\alpha$ photons and spin temperature
is proportional to the intensity $J$ (e.g. Furlanetto, Oh, \& Briggs
 2006). The W-F coupling would generally be suppressed due to the
fact that the flux at the resonant frequency $J(x=0)$ is always less
than the flux at other frequencies. In a saturated state this
suppression is small (Figure 6). However, before approaching the
saturated state, the intensity $J$ generally is significantly less
than its saturated value. That is, although the local Boltzmann
distribution is formed at the time of the order of $\tau\simeq
10^4$, the intensity at that time would still be low, and the W-F
coupling is not enough to produce the deviation of $T_s$ from
$T_{\rm CMB}$. Therefore, it may not be always reasonable to assume
that the deviation of $T_s$ from $T_{\rm CMB}$ is mainly due to the
W-F coupling if first stars or their emission/absorption regions
evolved with time scale equal to or less than Myrs.

The WENO algorithm revealed the time evolution of photons undergoing
resonant scattering, whose information is generally lost in the
asymptotic solutions, or the time-independent solution. Although
time-independent solutions provide useful guidance, they do not show
the conditions for the efficiency of the W-F coupling at different
times. The asymptotic solution probably is never reached for short
life-time objects. It would be impossible to correctly estimate
observable 21 cm signal from ionized and heated halos of first stars
without a correct understanding of the time evolution of the W-F
coupling.

\acknowledgments

This work is supported by the US NSF under the grants AST-0506734
and AST-0507340.

\appendix

\section{Numerical algorithm}

\subsection{Computational domain and computational mesh}

The computational domain in the case of static background is $x \in
[-15, 15]$. The initial condition is $J(x, 0)=0$ and the boundary condition is
$J(x, \tau) =0$ at the boundaries.
In the case of expanding background ($\gamma\neq0$), the
computational domain is bigger, depending on the
value of the Sobolev parameter $\gamma$. The domain, denoted as
$(x_{left},x_{right})$, is chosen such that
$J(x_{left}, \tau) \approx 0$ and $J(x_{right}, \tau) \approx 0$.
For example,  the domain is taken to be $x \in [-100,15]$ for the
case of $\gamma=10^{-3}$.

The computational domain $(x_{left},x_{right})$ is discretized into a
uniform mesh as following,
$$x_i=i\Delta x, \hspace{25mm}   i = -N_l \cdots,N_r,$$
where $N= N_l + N_r$ and $\Delta x=(x_{right}-x_{left})/N$, is the
mesh size. We also denote $J_i^n = J(x_i,\tau^n)$, the approximate
solution values at $x_i$ and the $n^{th}$ time step, i.e. $\tau^n =
n\Delta t$, $\Delta t$ being the numerical time step.

\subsection{The WENO algorithm: approximations to spatial derivatives}

To calculate $\frac{\partial J}{\partial x}$, we use the fifth order
WENO method (Jiang and Shu, 1996). That is,
\begin{equation}
\frac{\partial J(x_i,\tau^n)}{\partial x} \approx \frac{1}{\Delta x}
(\hat{h}_{i+1/2}-\hat{h}_{i-1/2}\\)
\end{equation}
where the numerical flux $\hat{h}_{i+1/2}$ is obtained by the
procedure given below. We use the upwind flux in the fifth order
WENO approximation because the wind direction is fixed (negative).
First, we denote
\begin{equation}
h_i = J(x_i,\tau^n),    \hspace{25mm} i = -2, -1,\cdots,N+3\\
\end{equation}
where $n$ is fixed. The numerical flux from the WENO procedure is
obtained by
\begin{equation}
\hat{h}_{i+1/2}=\omega_1\hat{h}_{i+1/2}^{(1)}+\omega_2\hat{h}_{i+1/2}^{(2)} +\omega_3\hat{h}_{i+1/2}^{(3)},\\
\end{equation}
where $\hat{h}_{i+1/2}^{(m)}$ are the three third order fluxes on
three different stencils given by
\begin{eqnarray*}
\hat{h}_{i+1/2}^{(1)} &=& -\frac{1}{6}h_{i-1}+\frac{5}{6}h_{i}+\frac{1}{3}h_{i+1},\\
\hat{h}_{i+1/2}^{(2)} &=& \frac{1}{3}h_{i}+\frac{5}{6}h_{i+1}-\frac{1}{6}h_{i+2},\\
\hat{h}_{i+1/2}^{(3)} &=&
\frac{11}{6}h_{i+1}-\frac{7}{6}h_{i+2}+\frac{1}{3}h_{i+3},
\end{eqnarray*}
and the nonlinear weights $\omega_m$ are given by,
\begin{equation}
\omega_m =
\frac{\check{\omega}_m}{\displaystyle\sum_{l=1}^3\check{\omega}_l},
\hspace{5mm}
\check{\omega}_l = \frac{\gamma_l}{(\epsilon+\beta_l)^2},\\
\end{equation}
where $\epsilon$ is a parameter to avoid the denominator to become
zero and is taken as $\epsilon = 10^{-8}$. The linear weights
$\gamma_l$ are given by
\begin{equation}
\gamma_{1} = \frac{3}{10},\hspace{3mm} \gamma_{2} = \frac{3}{5},
\hspace{3mm} \gamma_{3} = \frac{1}{10},
\end{equation}
and the smoothness indicators $\beta_{l}$ are given by,
\begin{eqnarray*}
\beta_1 &=& \frac{13}{12}(h_{i-1}-2h_{i}+h_{i+1})^2 +\frac{1}{4}(h_{i-1}-4h_{i}+3h_{i+1})^2,\\
\beta_2 &=& \frac{13}{12}(h_{i}-2h_{i+1}+h_{i+2})^2 +\frac{1}{4}(h_{i}-h_{i+2})^2,\\
\beta_3 &=& \frac{13}{12}(h_{i+1}-2h_{i+2}+h_{i+3})^2
+\frac{1}{4}(3h_{i+1}-4h_{i+2}+h_{i+3})^2.
\end{eqnarray*}

\subsection{Numerical integration: an $\mathcal{O}(N)$ algorithm}

We need to numerically integrate $\displaystyle\int \mathcal{R}(x, x') J(x', t) dx'$, denoted as
\beq
I(x) =  \frac12 \int  e^{2bx' + b^2} {\rm erfc} (\max(|x+b|, |x'+b|)) J(x', t) dx',
\eeq
with $\mathcal{R}(x, x')$ as in eq.(\ref{eq:R}).
To evaluate $I_m = I(x_m)$,  $\forall m=-N_l, \cdots, N_r$, we apply the rectangular rule,
which is spectrally accurate for smooth functions vanishing at boundaries,
\beqa
I_m
& =&  \frac 12 \int_{x_{left}}^{x_{right}} {\rm erfc} (\max(|x_m+b|, |x'+b|))e^{2bx' + b^2} J(x', t)dx' \\
&\approx&\frac12 \Delta x \sum_{i=-N_l}^{N_r} {\rm erfc} (\max(|x_m+b|, |x_i+b|))e^{2bx_i + b^2}J(x_i, t) .
\eeqa
Notice that this summation algorithm is very costly as it takes $\mathcal{O}(N)$
operations per $m$, therefore the total procedure has $\mathcal{O}(N^2)$ operations overall.
We use a grouping technique, described below, so that the overall computational cost
can be reduced to $\mathcal{O}(N)$, without changing mathematically the algorithm and its accuracy.

 The proposed scheme with order $N$ computational effort is the following.
Let $N_b = floor(\frac{b}{\Delta x})$ and $N_{2b} = floor(\frac{2b}{\Delta x})$.
The integration algorithm is designed for two cases:
$m \ge -N_b$ and $m< -N_b$.

In the case of $m \ge -N_b$ or equivalently $x_m + b \ge 0$:
\beqa
I_m  &=& \frac12 \Delta x  (\sum_{i=-N_l}^{-m-N_{2b}-1} {\rm erfc} (|x_i+b|)e^{2bx_i + b^2}J(x_i,
t)  \nonumber \\
& & +  {\rm erfc} (|x_m+b|) \sum_{i=-m - N_{2b}}^{m}e^{2bx_i + b^2}J(x_i, t) \nonumber\\
& & +\sum_{i=m+1}^{N_r} {\rm erfc} (|x_i+b|)e^{2bx_i + b^2}J(x_i, t))\\
&\doteq& \frac12 \Delta x (I_{1, m} + {\rm erfc} (|x_m+b|) I_{2, m} + I_{3,m})
\eeqa

\begin{enumerate}
\item Evaluate $I_{1, -N_b}$, $I_{2, -N_b}$ and $I_{3,-N_b}$ respectively as
\beqa
I_{1, -N_b} & = & \sum_{i=-N_l}^{N_b-N_{2b}-1} {\rm erfc} (|x_i+b|)e^{2bx_i + b^2}J(x_i, t), \\
I_{2, -N_b} &=& \sum_{i=N_b - N_{2b}}^{-N_b} e^{2bx_i + b^2}J(x_i, t), \\
I_{3,-N_b} &=& \sum_{i=-N_b+1}^{N_r} {\rm erfc} (|x_i+b|)e^{2bx_i + b^2}J(x_i, t),
\eeqa
which leads to $\mathcal{O}(N)$ cost.
\item
Do $m = -N_b+1, N_r$

Evaluate $I_{1, m}$, $I_{2, m}$, $I_{3, m}$ respectively by
\beq
I_{1, m} = I_{1, m-1} - {\rm erfc} (|x_{-m-N_{2b}}+b|) e^{2bx_{-m-N_{2b}}+ b^2}J(x_{-m-N_{2b}}, t)
\eeq
\beq
I_{2, m} = I_{2, m-1} + e^{2bx_{m} + b^2}J(x_{m}, t) + e^{2bx_{-m-N_{2b}} + b^2}J(x_{-m-N_{2b}}, t)
\eeq
\beq
I_{3, m} = I_{3, m-1} -{\rm erfc} (|x_{m}+b|)e^{2bx_{m} + b^2}J(x_{m}, t)
\eeq
ENDDO

To be consistent with the indexes, if $N_l - N_{2b} < N_r$ then, we will set $I_{1, m} = 0$, for $m = N_l - N_{2b}, N_r$. 
The algorithm leads to $\mathcal{O}(1)$ cost per $m$, therefore $\mathcal{O}(N)$ computation overall.
\end{enumerate}

In the case of $m < -N_b$, or equivalently $x_m + b < 0$:
\beqa
I_m  &=& \frac12 \Delta x  (\sum_{i=-N_l}^{m-1} {\rm erfc} (|x_i+b|)e^{2bx_i + b^2}J(x_i, t)  \nonumber\\
&&+  {\rm erfc} (|x_m+b|) \sum_{i=m}^{-m-N_{2b}-1}e^{2bx_i + b^2}J(x_i, t) \nonumber\\
& & +\sum_{i=-m-N_{2b}}^{N_r} {\rm erfc} (|x_i+b|)e^{2bx_i + b^2}J(x_i, t))\\
&=& \frac12 \Delta x (I_{1, m} + {\rm erfc} (|x_m+b|) I_{2, m} + I_{3,m})
\eeqa
\begin{enumerate}
\item Evaluate $I_{1, -N_b-1}$, $I_{2, -N_b-1}$ and $I_{3,-N_b-1}$ as
\beqa
I_{1, -N_b-1} &=& \sum_{i=-N_l}^{-N_b-2} {\rm erfc} (|x_i+b|)e^{2bx_i + b^2}J(x_i, t), \\
I_{2, -N_b-1} &=&\sum^{N_b - N_{2b}}_{i=-N_b-1} e^{2bx_i + b^2}J(x_i, t), \\
I_{3,-N_b-1} &=&\sum_{i=N_b+1-N_{2b}}^{N_r} {\rm erfc} (|x_i+b|)e^{2bx_i + b^2}J(x_i, t),
\eeqa
which leads to $\mathcal{O}(N)$ cost.
\item
Do $m = -N_b-2, -N_l$

Evaluate $I_{1, m}$, $I_{2, m}$, $I_{3, m}$ respectively by
\beq
I_{1, m} = I_{1, m+1} - {\rm erfc} (|x_{m}+b|) e^{2bx_{m}+ b^2}J(x_{m}, t)
\eeq
\beq
I_{2, m} = I_{2, m+1} + e^{2bx_{m} + b^2}J(x_{m}, t) + e^{2bx_{-m-N_{2b}-1} + b^2}J(x_{-m-N_{2b}-1}, t)
\eeq
\beq
I_{3, m} = I_{3, m+1} - {\rm erfc} (|x_{-m-N_{2b}-1}+b|)e^{2bx_{-m-N_{2b}-1} + b^2}J(x_{-m-N_{2b}-1}
, t)
\eeq
ENDDO

To be consistent with the indexes, if $N_l - N_{2b} > N_r$, we will set $I_{3, m} = 0$, for $m= -N_{2b} - N_r, -N_l$. 
Again, the algorithm leads to $\mathcal{O}(1)$ cost per $m$, therefore $\mathcal{O}(N)$ computation overall.
\end{enumerate}

\subsection{Time evolution}

To evolve in time, we use the third-order TVD Runge Kutta time
discretization (Shu \& Osher, 1988). For systems of ODEs $u_t =
L(u)$, the third order Runge-Kutta method is
\begin{eqnarray*}
u^{(1)} &=& u^n + \Delta \tau L(u^n,\tau^n),\\
u^{(2)} &=& \frac{3}{4}u^n + \frac{1}{4}(u^{(1)}+\Delta \tau
L(u^{(1)},\tau^n +
\Delta \tau)),\\
u^{n+1} &=& \frac{1}{3}u^n + \frac{2}{3}(u^{(2)}+\Delta \tau
L(u^{(2)},\tau^n + \frac12 \Delta \tau)).
\end{eqnarray*}


\end{document}